\documentclass[AMA,STIX2COL,Linenumberson]{MRM}
\articletype{PRE-PRINT - SUBMITTED TO MAGNETIC RESONANCE IN MEDICINE}%

\usepackage{defs}

\begin{document}

\title{A Review of Machine Learning Applications for the Proton Magnetic Resonance Spectroscopy Workflow}

\author[1]{Dennis M. J. van de Sande}{*\orcid{0000-0001-6112-1437}}
\author[2]{Julian P. Merkofer}{*\orcid{0000-0003-2924-5055}}
\author[1]{Sina Amirrajab}{\orcid{0000-0001-8226-7777}}
\author[1]{Mitko Veta}{\orcid{0000-0003-1711-3098}}
\author[2,3]{Ruud J. G. van Sloun}{\orcid{0000-0003-2845-0495}}
\author[4]{Maarten J. Versluis}{\orcid{0000-0003-3601-0377}}
\author[2,5,6]{Jacobus F. A. Jansen}{\orcid{0000-0002-5271-8060}}
\author[4]{Johan S. van den Brink}{\orcid{0000-0002-7058-4449}}
\author[1,2,4]{Marcel Breeuwer}{\orcid{0000-0003-1822-8970}}

\authormark{}

\address[1]{\orgdiv{Department of Biomedical Engineering}, \orgname{Eindhoven University of Technology}, \orgaddress{\state{Eindhoven}, \country{The Netherlands}}}

\address[2]{\orgdiv{Department of Electrical Engineering}, \orgname{Eindhoven University of Technology}, \orgaddress{\state{Eindhoven}, \country{The Netherlands}}}

\address[3]{\orgdiv{Philips Research}, \orgname{Philips Research}, \orgaddress{\state{Eindhoven}, \country{The Netherlands}}}

\address[4]{\orgdiv{MR R\&D - Clinical Science}, \orgname{Philips Healthcare}, \orgaddress{\state{Best}, \country{The Netherlands}}}

\address[5]{\orgdiv{Department of Radiology and Nuclear Medicine}, \orgname{Maastricht University Medical Center}, \orgaddress{\state{Maastricht}, \country{The Netherlands}}}

\address[6]{\orgdiv{School for Mental Health and Neuroscience}, \orgname{Maastricht University}, \orgaddress{\state{Maastricht}, \country{The Netherlands}}}

\corres{Dennis M. J. van de Sande 
{\hfill\break}Eindhoven University of Technology 
{\hfill\break}PO Box 513, 5600 MB Eindhoven, 
{\hfill\break}The Netherlands
{\hfill\break}\email{d.m.j.v.d.sande@tue.nl}}


\finfo{This work was (partially) funded by Spectralligence (EUREKA IA Call, ITEA4 project 20209).}

\abstract[Abstract]{This literature review presents a comprehensive overview of \acf{ml} applications in proton \acf{mrs}. As the use of \ac{ml} techniques in \ac{mrs} continues to grow, this review aims to provide the \ac{mrs} community with a structured overview of the state-of-the-art methods. Specifically, we examine and summarize studies published between 2017 and 2023 from major journals in the magnetic resonance field. We categorize these studies based on a typical \ac{mrs} workflow, including data acquisition, processing, analysis, and artificial data generation. Our review reveals that \ac{ml} in \ac{mrs} is still in its early stages, with a primary focus on processing and analysis techniques, and less attention given to data acquisition. We also found that many studies use similar model architectures, with little comparison to alternative architectures. Additionally, the generation of artificial data is a crucial topic, with no consistent method for its generation. Furthermore, many studies demonstrate that artificial data suffers from generalization issues when tested on in-vivo data. We also conclude that risks related to \ac{ml} models should be addressed, particularly for clinical applications. Therefore, output uncertainty measures and model biases are critical to investigate. Nonetheless, the rapid development of \ac{ml} in \ac{mrs} and the promising results from the reviewed studies justify further research in this field.}


\maketitle

\footnotetext{*Equally contributing authors}

\section{Introduction}\label{sec1}
\Acf{mrs} and \ac{mrsi} are non-invasive methods for investigating the chemical and structural properties of molecules in-vivo. These techniques are widely used for measuring human metabolism, particularly in the areas of neural diseases, tumor detection, and monitoring \cite{buonocore_magnetic_2015, faghihi_magnetic_2017, wilson_methodological_2019}. While \ac{mrs} and \ac{mrsi} have the potential to be highly valuable in clinical practice, they pose several challenges such as low \ac{snr}, overlapping metabolite signals, experimental artifacts, and long acquisition times. To effectively analyze spectroscopy data, various considerations such as pulse sequence selection \cite{landheer_theoretical_2020}, B0 shimming~\cite{juchem_b0_2021}, as well as preprocessing and analysis methods \cite{jansen_1h_2006, near_preprocessing_2021} must be taken into account. Due to the complexity of these considerations, \ac{mrs} and \ac{mrsi} can be challenging techniques for non-experts to implement and oversee, hindering clinical adoption \cite{wilson_methodological_2019}. 

The ability to learn model-agnostic features from data has made \acf{ml} methods very popular in many disciplines over the last decade. In \ac{mri} the use of \ac{ml} techniques, for example, has increasingly found a wide range of applications ranging from image reconstruction~\cite{knoll_deep-learning_2020, montalt-tordera_machine_2021, pal_review_2022} and quality improvement~\cite{zhu_applications_2019} to image analysis \cite{mohammed_review_2020} and clinical diagnostics \cite{uddin_progress_2017, rashid_towards_2020, pilmeyer_functional_2022, santana_rs-fmri_2022}. This trend has started to increase in \ac{mrs} and \ac{mrsi} as well, with various \ac{ml} methods being proposed to address some of the associated challenges. In the work of Chen et al.~\cite{chen_review_2020} a sparse collection of such \ac{dl}-based approaches is summarized. The work covers nine application examples in the domains of spectral reconstruction and denoising of proton \ac{mrs} as well as chemical shift prediction and automated peak-picking for proton and other \ac{nmr} spectroscopy. Another review by Rajeev et al.~\cite{rajeev_review_2021} focuses on the clinical diagnosis of brain tumors from \ac{mr} spectra using \ac{dl} methods. The study condenses twenty data-driven approaches designed to improve the \ac{mrs} workflow and consequently improve tumor diagnosis. 
However, an exhaustive collection of recent \ac{ml} applications in \ac{mrs} is still missing. Furthermore, these previous reviews do not show where the discussed \ac{ml} studies fit into the \ac{mrs} workflow, which inhibits better insight into the application domain. Moreover, with continuously emerging techniques in \ac{ml}~\cite{wang_recent_2020, marcus_next_2020, sarker_deep_2021}, the urgency for a thorough documentation of \ac{ml} developments in \ac{mrs} has grown persistently.

In this review we aim to bridge the gap between specific knowledge of the \ac{mrs} workflow, from acquisition to clinical applications, and the technicalities of \ac{ml} methods. Comprehensive and assessable summaries of recent \ac{ml} studies are provided, based on their organization within common workflows of proton \ac{mrs}. We discuss and summarize architectures, input and output schemes, training strategies, and the intended application for a selection of studies.

\subsection{Literature Search}
\begin{figure}
\centerline{\includegraphics[width=\columnwidth]{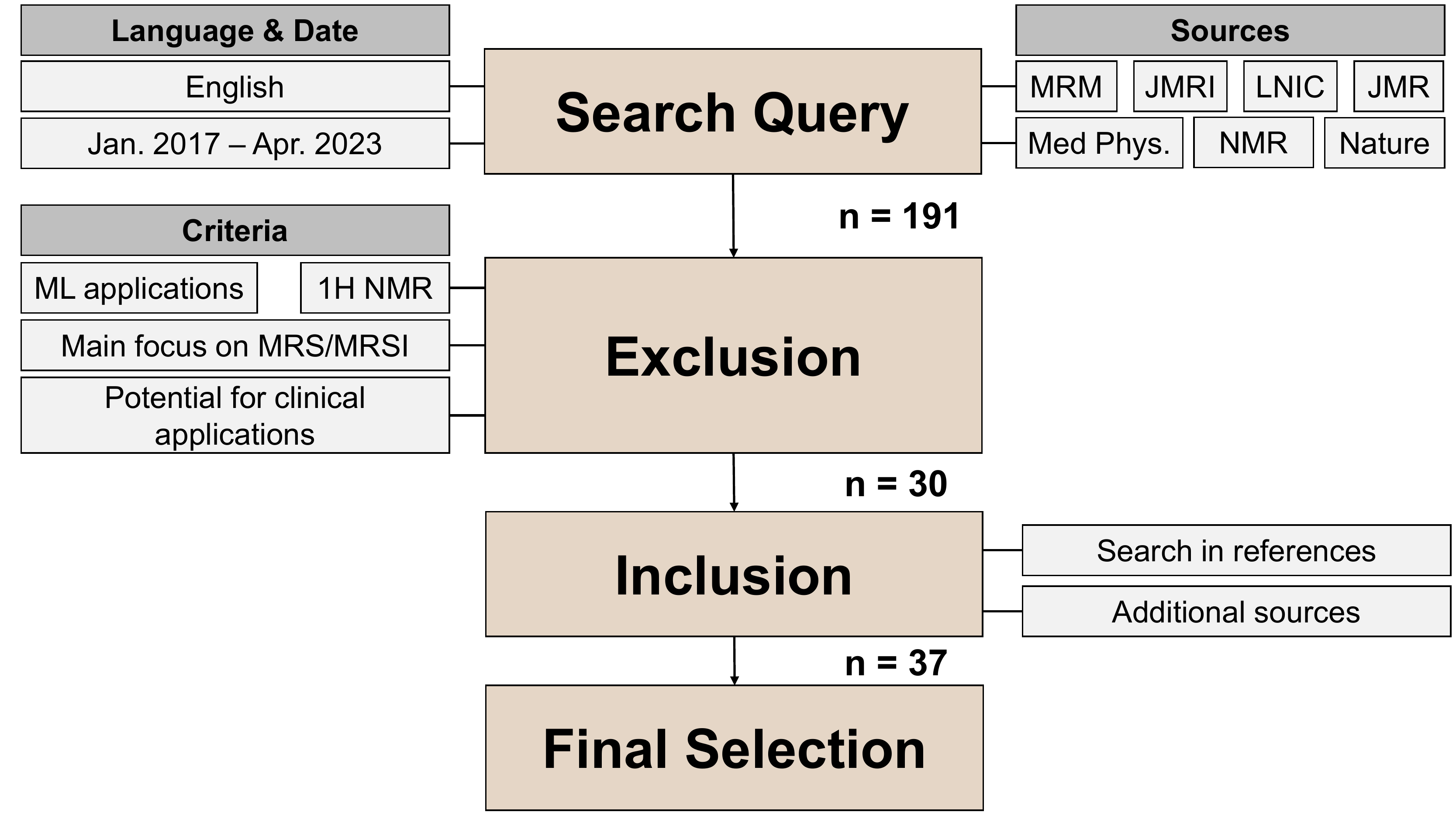}}
    \caption{An outline of the systematic literature search. At start, the language and time period are set before a filter is applied for the selected sources. The Scopus search query results in a selection of publications which are manually checked on the exclusion criteria. After including additional sources, the final selection is obtained. The amount of papers after each step is indicated with $n$.}
    \label{fig:scope}
\end{figure}

The literature search is conducted based on the systematic process outlined in Figure \ref{fig:scope}. To ensure a comprehensive overview, we focus on state-of-the-art \ac{ml} studies of the last seven years, as they hold the most relevance for current developments within the field. Using Elsevier's Scopus database the search is narrowed to studies published between and including January 2017 and April 2023 in major journals in the field of \ac{mr}. By determining a specific query to search in title, keywords, and abstract for specific keywords related to \ac{mrs}, \ac{mrsi}, \ac{ml}, \ac{dl}, and \acp{nn} the search is further limited to 191 studies. Literature is excluded if it primarily focuses on other modalities than \ac{mrs} and \ac{mrsi} or do not mention \ac{ml} applications. The final selection is obtained after investigating the references of the found literature and a final search using other search engines. Additional literature from other sources is added if their content fits within the scope. Table \ref{tab:overview_papers} provides an overview of the final 37 papers, covered in this review.

\begin{sidewaystable*}
\centering
\setlength{\tabcolsep}{3pt}
\caption{Overview of all summarized and discussed studies with information about year, workflow category, training data type, model type, input, and output.\label{tab:overview_papers}}%
\begin{tabular*}{\textwidth}{@{\extracolsep\fill}lllllll@{\extracolsep\fill}}%
\toprule
  \textbf{Study} & \textbf{Year} & \textbf{Category} & \textbf{Training Data Type} & \textbf{Model Type} & \textbf{Input} & \textbf{Output} \\
\midrule
Bolan et al. \cite{bolan_automated_2020}                           & 2020        & Volume Selection                         & In-Vivo                & U-Net       & T2w-FLAIR              &  Segmentation Mask    \\
Becker et al. \cite{becker_deep_2022}                              & 2022        & Shimming                                 & Phantom               & Ensemble    & 4D Spectra              &  Shim Values          \\
Becker et al. \cite{becker_acquisitions_2022}                      & 2022        & Shimming                                 & Phantom               & LSTM        & Spectrum + Shim Offset &  Shim Values          \\
Nassirour et al. \cite{nassirpour_multinet_2018}                   & 2018        & Reconstruction                           & In-Vivo                & Multiple Single-Layer NNs &  Undersampled k-Space  & k-Space Values  \\
Lee et al. \cite{lee_reconstruction_2020}                          & 2020        & Reconstruction                           & In-Vivo + Artificial   & U-Net  & FID/Spectrum & FID/Spectrum \\
Luo et al. \cite{luo_fast_2020}                                    & 2020        & Reconstruction                           & Artificial             & HRNet  & NUS L-COSY Spectrum & L-COSY Spectrum \\
Iqbal et al. \cite{iqbal_deep_2021}                                & 2021        & Reconstruction \& Quantification          & Artificial             & U-Net   & NUS L-COSY Spectrum & L-COSY Spectrum \\
Motyka et al. \cite{motyka_kspacebased_2021}                       & 2021        & Reconstruction                           & In-Vivo                & Shallow-Graph CNN & Encoding Step + FID Point & k-Space Points \\
Chan et al. \cite{chan_improved_2022}                              & 2022        & Reconstruction                           & In-Vivo                & Multiple Single-Layer NNs &  Undersampled k-Space  & k-Space Values  \\
Lei et al. \cite{lei_deep_2021}                                    & 2021        & Spectral Denoising                       & In-Vivo + Phantom      & Autoencoder  & Low NSA Spectrum  & High NSA Spectrum \\
Iqbal et al. \cite{iqbal_super-resolution_2019}                    & 2019        & Super-Resolution MRSI                          & In-Vivo + Artificial   & U-Net  &  LRSI Image    & HRSI Image     \\
Dong et al. \cite{dong_high-resolution_2021}                       & 2021        & Super-Resolution MRSI                          & In-Vivo                    & U-Net  &  LRSI Image    & HRSI Image     \\
Tapper et al. \cite{tapper_frequency_2021}                         & 2021        & Frequency \& Phase Correction           & Artificial                 & MLP    &  Spectrum  & Frequency/Phase Values \\
Ma et al. \cite{ma_mr_2022}                                        & 2022        & Frequency \& Phase Correction           & Artificial                 & CNN    &  Spectrum  & Frequency/Phase Values \\
Shamaei et al. \cite{shamaei_modelinformed_2022}                   & 2022        & Frequency \& Phase Correction           & Artificial                 & Autoencoder &  FID         & FID (Corrected)         \\
Kyathanahally et al. \cite{kyathanahally_deep_2018}                & 2018        & Ghosting Artifact Removal   & Artificial  & MLP, CNN, Autoencoder & Spectrum/Spectrogram  & Ghost Class (2)/Spectrogram \\
Lee and Kim \cite{lee_intact_2019}                                 & 2019        & General Artifact Removal                 & Artificial                 & CNN  & Spectrum    & Spectrum (Metabolites Only)      \\
Pedrosa de Barros et al. \cite{pedrosa_de_barros_improving_2017}   & 2017        & Quality Assurance                        & In-Vivo            & Random Forest  & FID + Spectral Features  & Quality Class (2) \\
Gurbani et al. \cite{gurbani_convolutional_2018}                   & 2018        & Quality Assurance                        & In-Vivo                    & CNN  &  Spectrum                  & Quality Class (3)  \\
Kyathanahally et al. \cite{kyathanahally_quality_2018}             & 2018        & Quality Assurance                        & In-Vivo                & SVM, LDA, RUSBoost & Spectral Features  & Quality Class (3) \\
Jang et al. \cite{jang_unsupervised_2021}                          & 2021        & Quality Assurance                        & Artificial                 & GAN &  Spectrum                 & Quality Class (2) \\
Hernández-Villegas et al. \cite{hernandez-villegas_extraction_2022}& 2022        & Quality Assurance                        & In-Vivo                    & NNMF & Spectrum            & Quality Class (3) \\
Das et al. \cite{das_quantification_2017}                          & 2017        & Quantification                           & In-Vivo + Artificial    & Random Forest & Spectrum          & Metabolite Concentrations     \\
Hatami et al. \cite{hatami_magnetic_2018}                          & 2018        & Quantification                           & Artificial                 & CNN  & Spectrum        & Metabolite Concentrations  \\
Gurbani et al. \cite{gurbani_incorporation_2019}                   & 2019        & Quantification                           & In-Vivo                    & Autoencoder  & Spectrum   & Spectrum          \\
Lee and Kim \cite{lee_deep_2020}                                   & 2020        & Quantification \& Uncertainty Meas.      & Artificial + Phantom       & CNN & Spectrum           & Spectrum (Metabolite Only)             \\
Shamaei et al. \cite{shamaei_wavelet_2021}                         & 2021        & Quantification                           & Artificial                 & CNN & FID                   & Metabolite Concentrations \\
Rizzo et al. \cite{rizzo_quantification_2023}                      & 2023        & Quantification                           & Artificial                 & CNN, Ensemble & Spectrum/Spectrogram   & Metabolite Concentrations \\
Schmid et al. \cite{schmid_deconvolution_2023}                     & 2023        & Quantification                           & Artificial                 & CNN, LSTM & Spectrum                  & Peak Class (3)/Peak Widths \\
Shamaei et al. \cite{shamaei_physics-informed_2023}                & 2023        & Quantification                           & In-Vivo + Artificial       & Autoencoder & FID                  & FID \\
Lee and Kim \cite{lee_bayesian_2022}                               & 2022        & Uncertainty Measurement                  & Artificial                 & CNN & Spectrum                       & Spectrum          \\
Rizzo et al. \cite{wang_reliability_2022}                          & 2022        & Uncertainty Measurement                  & Artificial                 & CNN & Spectrogram                & Metabolite Concentrations \\
Zarinabad et al. \cite{zarinabad_multiclass_2017}                  & 2017        & Classification                           & In-Vivo                & Ensemble & Spectrum/Metabolite Concentrations & Tumor Class (3) \\
Zarinabad et al. \cite{zarinabad_application_2018}                 & 2018        & Classification                           & In-Vivo                    & SVM, LDA, Random Forrest & Metabolite Concentration Features & Tumor Class (3) \\
Dikaios \cite{dikaios_deep_2021}                                   & 2021        & Classification                           & In-Vivo + Artificial    & SVM, MLP, CNN & Spectrum        & Tumor Class (2)       \\
Zhao et al. \cite{zhao_metabolite_2022}                            & 2022        & Classification                           & In-Vivo                    & SVM, LDA, k-Means, Naive Bayes, NN & Metabolite Concentration Features & Tumor Class (3) \\
Olliverre et al. \cite{olliverre_generating_2018}                  & 2018        & ML-Based Artificial Data Generation      & In-Vivo                 & GAN       & Noise Vector            & Spectrum            \\             
\bottomrule
\end{tabular*}

\end{sidewaystable*}

\subsection{MRS Workflow}
This review is structured following a common \ac{mrs} and \ac{mrsi} workflow ~\cite{near_preprocessing_2021, kreis_terminology_2021}. This workflow is applicable for clinical and research purposes and is divided into three main parts: data acquisition, processing, and analysis. 

\begin{figure}[h]
    \centering
    \includegraphics[width=\columnwidth]{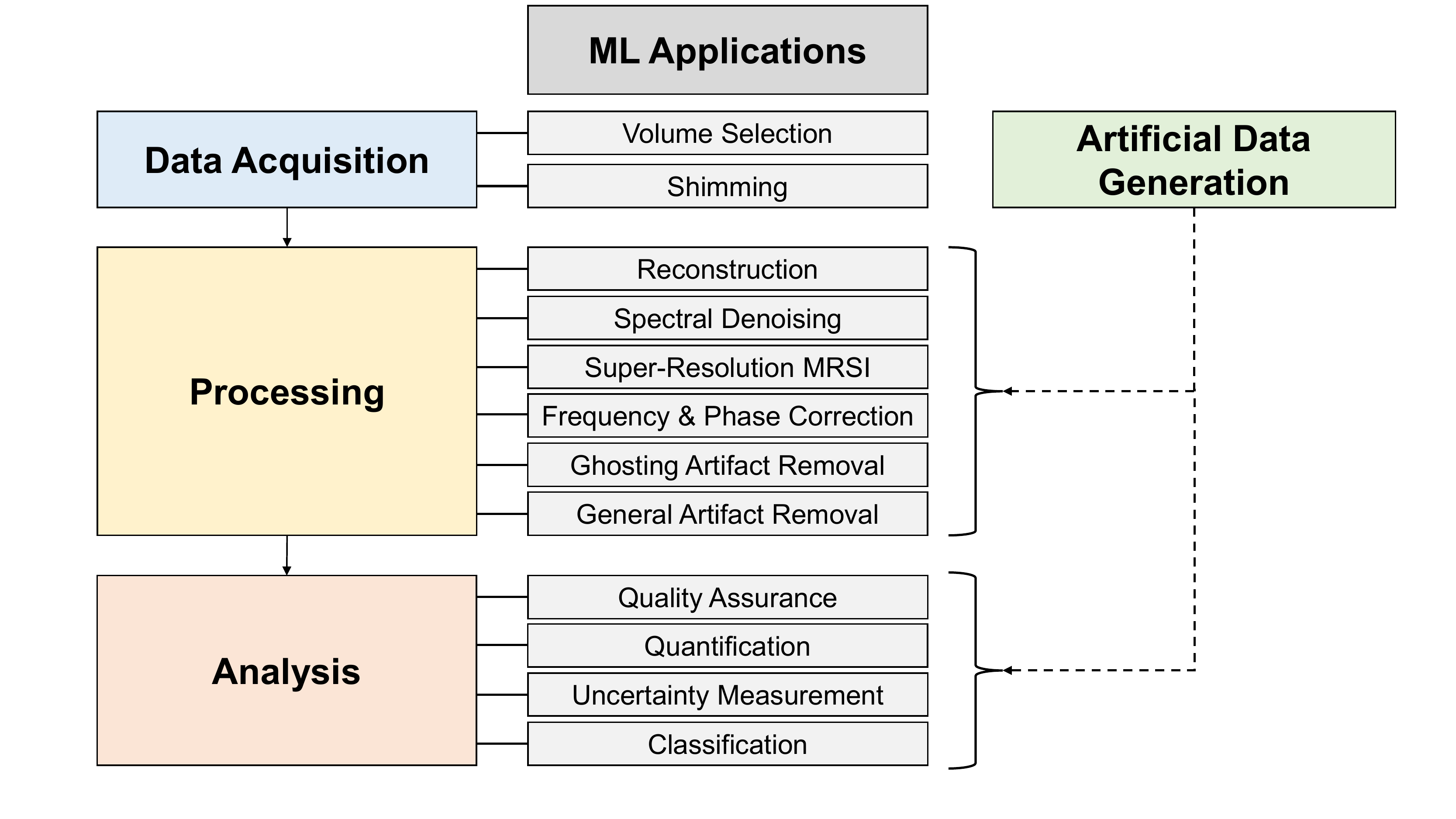}
    \caption{Schematic overview of the overall \ac{mrs} and \ac{mrsi} workflow with corresponding \ac{ml} applications. The dashed arrows indicate a possibility to include artificial data into the development of some \ac{ml} applications.}
    \label{fig:pipeline}
\end{figure}

Data acquisition includes all the necessary steps for acquiring raw \ac{mrs} or \ac{mrsi} data, such as pulse sequence design, voxel placement, and $B_0$ shimming. The processing step involves techniques that reduce the dimensionality of the data, remove spectral imperfections, or improve the visual appearance of the spectra. Some examples include signal averaging, eddy current correction, residual water/lipid removal, motion correction, apodization, and zero-filling. The analysis step involves using the processed data to evaluate its quality, quantify it with uncertainty, or classify it by specific characteristics such as disease. \Ac{ml} can be applied at each step of this workflow to perform or improve specific tasks. Additionally, some \ac{ml} applications may require the use of artificial data for training, because there is a lack of large open databases. Since \ac{ml} methods are highly dependent on the training data, artificial data generation is added as a workflow category. Figure~\ref{fig:pipeline} provides a schematic overview of the workflow, highlighting the use of \ac{ml} at each step.

\begin{figure*}[h!]
    \centering 
\begin{minipage}[t]{.45\textwidth}
\begin{subfigure}{\textwidth}
  \includegraphics[width=\linewidth, page=1]{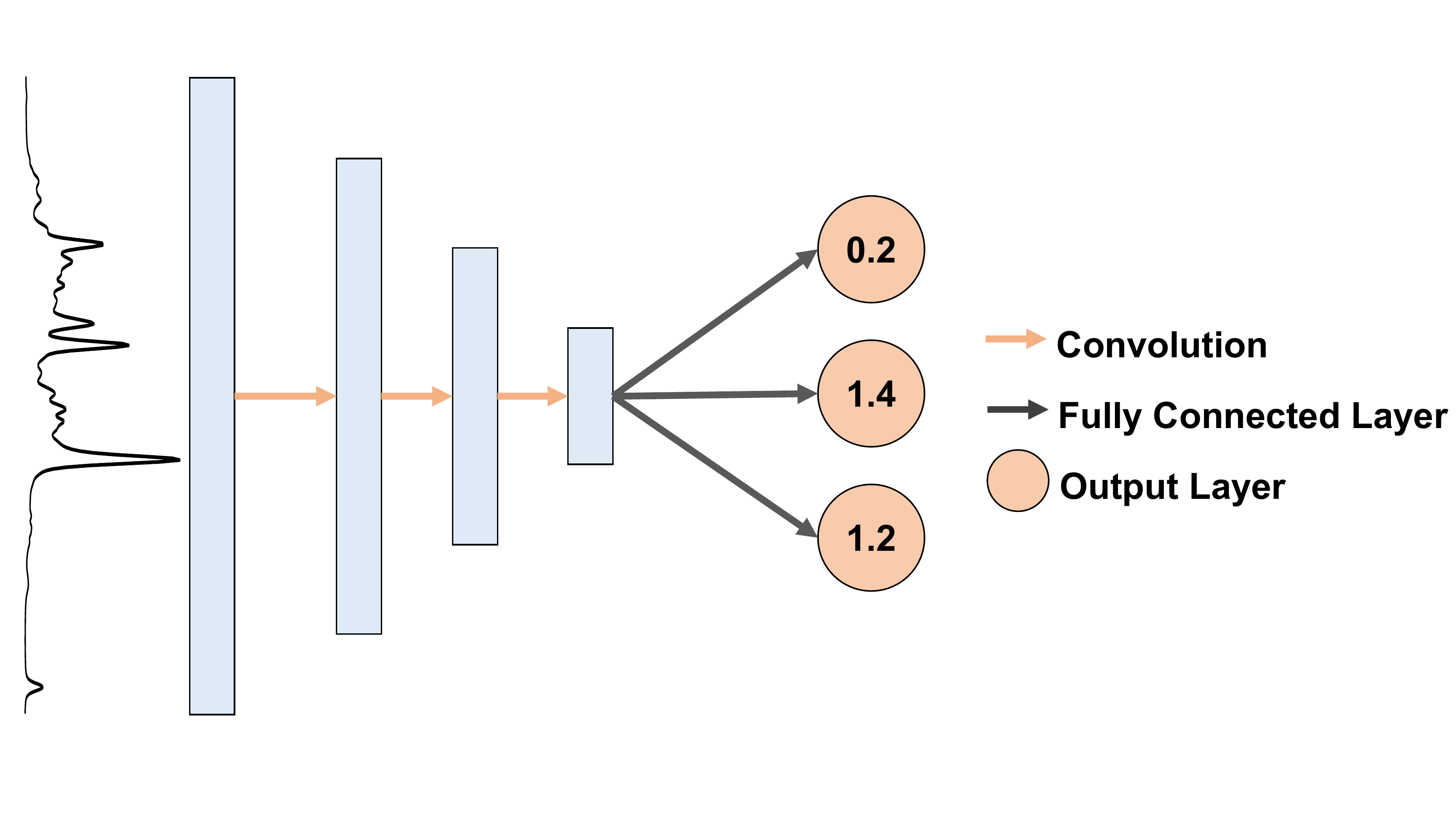}
  \caption{\Acf{cnn} model type.}
  \label{fig:1}
\end{subfigure}\hfil
\begin{subfigure}{\textwidth}
  \includegraphics[width=\linewidth, page=2]{model_types_v3.pdf}
  \caption{\Acf{mlp} model type.}
  \label{fig:2}
\end{subfigure}\hfil
\begin{subfigure}{\textwidth}
  \includegraphics[width=\linewidth, page=3]{model_types_v3.pdf}
  \caption{U-Net model type.}
  \label{fig:3}
\end{subfigure}
\end{minipage}\hfil
\begin{minipage}[t]{.45\textwidth}
\begin{subfigure}{\textwidth}
  \includegraphics[width=\linewidth, page=4]{model_types_v3.pdf}
  \caption{Autoencoder model type.}
  \label{fig:4}
\end{subfigure}\hfil
\begin{subfigure}{\textwidth}
  \includegraphics[width=\linewidth, page=5]{model_types_v3.pdf}
  \caption{\Acf{lstm} model type.}
  \label{fig:5}
\end{subfigure}\hfil
\begin{subfigure}{\textwidth}
  \includegraphics[width=\linewidth, page=6]{model_types_v3.pdf}
  \caption{\Acf{gan} model type.}
  \label{fig:6}
\end{subfigure}
\end{minipage}
\caption{Overview of the most commonly used model types that are discussed in this review. In \ref{fig:1} a \ac{cnn} architecture is visualized which takes a spectrum as input and outputs scalar values (i.e. metabolite concentrations). In \ref{fig:2} a \ac{mlp} is visualized with an arbitrary number of nodes. In \ref{fig:3} a U-Net model is visualized which performs super-resolution with a \ac{lrsi} image as input and a \ac{hrsi} image as output. In \ref{fig:4} a general autoencoder is visualized which aims to reconstruct the input spectrum in the output. In \ref{fig:5} a \ac{lstm} model is visualized which takes a (time) sequence input and uses feedback connections to calculate the next hidden layer. Variables $x_t$ and $h_t$ indicate the sequence value and hidden state at timestep $t$ respectively. In \ref{fig:6} a \ac{gan} model is visualized that generates artificial spectra using a generator and includes a discriminator to determine whether a spectrum is real or fake.}
\label{fig:model_types}
\end{figure*}

This review does not contain any introduction to \ac{ml} methodology. For a comprehensive overview of \ac{ml}, \ac{dl}, and general \ac{ai} techniques we refer the reader to the alternative sources \cite{lecun_deep_2015, goodfellow_deep_2016, burkov_hundred-page_2019}. Schematic examples of some \ac{dl} model types that are seen in Table \ref{tab:overview_papers}, are shown in Figure \ref{fig:model_types}. Additionally, we refer to alternative sources for principles and explanations of \ac{mrs} concepts\cite{kreis_terminology_2021, de_graaf_vivo_2019}. 

The structure of this review is as follows: in Section \ref{sec:data_acquisition}, relevant \ac{ml} studies on data acquisition in the context of \ac{mrs} and \ac{mrsi} are summarized and discussed. In Sections \ref{sec:processing} and \ref{sec:analysis} the same is done for processing and analysis respectively. Section \ref{sec:artificial_dat_gen} briefly discusses artificial data generation and in Section \ref{sec:conclusion_outlook} an overall conclusion and outlook on the use of \ac{ml} in \ac{mrs} and \ac{mrsi} is provided.

\section{Data Acquisition}\label{sec:data_acquisition}
During data acquisition, scan-configuration parameters need to be optimized to get the desired and optimal data output. In this section some \ac{ml}-based data acquisition methods are summarized and discussed. 

\subsection{Volume Selection}
Voxel placement in \ac{svs} is critical to limit partial volume effects, especially for the analysis of brain tumors. Bolan et al. \cite{bolan_automated_2020} propose an algorithm for automated voxel placement. They use a dataset of 60 low grade glioma patients containing T2w \ac{flair} images and corresponding \ac{mrs} voxels manually placed by an expert spectroscopist. Lesion masks are retrospectively annotated by an experienced neurooncologist to have a gold-standard for training. The first step in their method involves tumor segmentation using a pre-trained \ac{cnn} model that is fine-tuned with their own dataset. The obtained segmentation volumes are used to maximize an objective function that describes the placement of a cuboid voxel in terms of position, size and rotation angle. This function captures two main considerations: voxel size and lesion fraction. Evaluation is done by comparing the lesion fraction, the volume of intersection between the annotated lesion and the \ac{mrs} voxel, and the total voxel size between manually and automatically placed voxels. The authors found that the proposed automatic placement method has a higher lesion fraction compared to manually placed voxels. Moreover, their method demonstrates more consistent placement with lower standard deviations for lesion fraction, volume of intersection, and voxel size.

\subsection{Shimming}
Performing $B_0$ shimming is important to obtain useful and high-quality \ac{mrs} data \cite{juchem_b0_2021}. To accelerate and automate the shimming process, Becker et al. \cite{becker_deep_2022} proposes a \ac{dl}-based method for shimming. The used dataset contains raw $^1$H-\acs{fid} signals with shim offsets, in which only linear shims in orthogonal $X, Y$ and $Z$ directions are considered. The \ac{dl} method aims to predict shim values for the $X, Y$, and $Z$ directions based on the \ac{1d} spectra of the linear shim offsets. They use an ensemble model architecture consisting of heterogeneous weak learners that are combined by either averaging, a fully connected layer, or a \ac{mlp}. Results show that both a single weak learner and the ensemble model with a \ac{mlp} are able to predict shim values that improve spectral quality. These models are also used in combination with the downhill simplex method~\cite{press_numerical_2007}, which is well-established for automatic shimming. They found that using their models, as stand-alone or in combination with this simplex method, results in either a reduction in the number of acquisitions necessary or an improvement in spectral quality.

In a follow-up study, Becker et al. \cite{becker_acquisitions_2022} extend their previous work by incorporating a higher order shim ($Z^2$) and a different \ac{nn} architecture in their study. The proposed architecture uses a \ac{cnn} combined with a \ac{lstm} block in order to mimic a signal-based shimming technique where the previously obtained states, in the form of a \ac{1d} spectrum and a shim offset, are used in the process. The value of the shim offsets are dependent on the time step during training. First, a number of random steps are taken followed by a series of predictive steps for which the input shim offsets are the previously obtained values. The results show that using \ac{dl} methods, with or without traditional optimization algorithms, is more effective than using traditional optimization alone.

\section{Processing}\label{sec:processing}
Raw \ac{mrs} and \ac{mrsi} measurements require a multitude of processing steps to obtain interpretable signals. Commonly recommended steps for \ac{mrs} include coil combination, signal averaging, motion correction, eddy current correction, \ac{fpc}, and identification of spurious echoes \cite{near_preprocessing_2021}.The following sections present and summarize \ac{ml} studies that have shown to improve and replace such existing processing methods. 

\subsection{Reconstruction}
Efficient sampling and reconstruction techniques play an important role in accelerating \ac{mrs} and \ac{mrsi} methods. When the data is highly under-sampled, reconstructing spectra from truncated \acp{fid} creates truncation artifacts. Lee et al. \cite{lee_reconstruction_2020} propose and compare three different reconstruction approaches with identically designed U-Net \cite{ronneberger_u-net_2015} architectures. The networks differ based on their inputs and outputs, which are either completely in the time domain, completely in the frequency domain, or mixed (\ac{fid} in, spectrum out). In-vivo 9.4T rat brain spectra are acquired to test the approaches as well as extract knowledge of the present \ac{snr} and linewidth values. Training data is obtained using a basis set simulation for 17 metabolites. The results show that the U-Net operating purely in the frequency domain has the best performance in terms of lowest \ac{nmse} and highest Pearson correlation coefficient (for the simulated data). The following observations are made with the simulated data: for 8 and 16 retained points (out of 1024) the U-Net recovers spectra with substantial truncation artifacts; for 32 and 64 retained points the approach manages to recover spectra with minor residuals; for 128 (and upwards) the truncation artifacts are well suppressed by the \ac{nn}, enabling precise quantification.
 
In addition to \ac{1d} \ac{svs} there are \ac{2d} \ac{mrs} techniques, such as the \ac{lcosy} experiment~\cite{thomas_localized_2001}. Despite the long acquisition time for \ac{lcosy}, it can aid in distinguishing overlapping metabolites. Luo et al. \cite{luo_fast_2020} introduce an encoder-decoder network architecture for fast reconstruction of \ac{nus} \ac{lcosy} spectra by learning to predict fully-sampled spectra from under-sampled input spectra. They propose to use simulated training data generated by the mode of virtual echo~\cite{mayzel_causality_2014}, where the under-sampled spectra are obtained by exponential and Poisson-gap sampling. This method is found to have fewer reconstruction artifacts and better peak preservation compared to other architectures such as \ac{cnn} and U-Net. It is also further evaluated with multi-nuclei spectra and found to have similar reconstruction quality compared to an iterative soft thresholding approach \cite{hyberts_application_2012} as well as \ac{smile} reconstruction~\cite{ying_sparse_2017}.

To accelerate \ac{lcosy} experiments, Iqbal et al. \cite{iqbal_deep_2021} propose a U-Net model to reconstruct fully-sampled spectra using \ac{nus} spectra as an input. They test their approach on simulated \ac{lcosy} spectra with exponential sampling. Results suggest that the U-Net architecture not only produces good quality spectra for all tested acceleration factors (1.3$\times$, 2$\times$, and 4$\times$), but also outperforms the compressed-sensing results of an $L1$-norm minimization method for the higher acceleration factors. 

Various acceleration techniques have been explored for \ac{mrsi} \cite{vidya_shankar_fast_2019, bogner_accelerated_2021}, mainly through under-sampling of the k-space and reconstruction using compressed-sensing or parallel imaging techniques \cite{pruessmann_sense_1999, griswold_generalized_2002, breuer_controlled_2006}. Nassirpour et al.~\cite{nassirpour_multinet_2018} propose to improve the conventional \ac{grappa} reconstruction~\cite{griswold_generalized_2002} by estimating k-space weightings with multiple \acp{nn} to reduce lipid aliasing. They train two types of single-layer \acp{nn} to predict the missing data points; one for cross-neighbors and one for adjacent neighbors. During reconstruction, the two network types are deployed sequentially in such a way that first the cross-neighbor \acp{nn} fill in the missing k-space values and then the adjacent-neighbor \acp{nn} estimate the remaining points. Additionally, the authors propose variable density under-sampling schemes to achieve even higher acceleration factors and alter their \ac{ml} framework by first using 2-voxel cross-neighbor and adjacent-neighbor \acp{nn} before using the previously mentioned 1-voxel neighbor \acp{nn}. Although in this work the networks are trained in a subject-specific manner, various strategies are investigated in ~\cite{chan_improved_2022} to improve this approach with more samples. The results suggest the use of \acp{nn} for \ac{grappa} reconstruction reduces aliasing artifacts thereby positively impacting metabolite concentration maps and significantly boosting the performance compared to regular \ac{grappa}. 

Furthermore, Motyka et al. \cite{motyka_kspacebased_2021} proposes a k-space-based coil combination using geometric \ac{dl} to reduce the amount of processing data immediately after the acquisition instead of reducing the data in the image domain. Their approach utilizes a shallow-graph \ac{nn} \cite{zhou_graph_2020} to learn the k-space representation of the \ac{mrsi} data before the summation step of the coil combination. In-vivo data are augmented for training and pairs consisting of input and desired output for each partition encoding step and each \ac{fid} point are created. Additionally, white Gaussian noise is added to the training samples to increase the robustness of the method. The proposed method is compared to the conventional image-based coil combination iMUSICAL~\cite{moser_noncartesian_2019}. For both approaches, the metabolite concentration maps are similar to the \acp{crlbp} of LCModel~\cite{provencher_estimation_1993}. The proposed method performs comparable to iMUSICAL when evaluated for different \acp{snr} levels, slightly under-performing for high \ac{snr} domains.

\subsection{Spectral Denoising}
\Ac{svs} acquisition time linearly depends on the number of signal averages which are obtained to enhance the \ac{snr}. Learning the mapping between low \ac{nsa} spectra to high \ac{nsa} spectra can effectively denoise and improve the \ac{snr} of the \ac{mrs} signal.

Lei et al. \cite{lei_deep_2021} propose an autoencoder model to denoise \ac{mrs} spectra. For that purpose they acquire multiple phantom and in-vivo spectra with a \ac{nsa} of 192 and of 8, representing high and low \ac{snr}. The proposed network consists of an encoder-decoder architecture, taking the processed low \ac{nsa} spectra as the input in the frequency domain and outputting high \ac{nsa} spectra with reduced noise. To enlarge the data variability a patch-based input is used in combination with data augmentation. The network is optimized using the \ac{mse} of the output spectra and the \ac{gt} high \ac{snr} spectra including a $L1$-norm on the hidden feature vector to enforce sparsity. \Ac{snr} estimates of the input and output spectra show an \ac{snr} improvement of 40\% and 47\% for phantom and in-vivo data, respectively, showing potential to accelerate \ac{mrs} acquisitions by acquiring low \ac{nsa} spectra while maintaining spectral quality.

\subsection{Super-Resolution MRSI}
Long acquisition times and high field strengths are often necessary to obtain \acf{hrsi} data. Advanced post-processing methods to increase the resolution for \acf{lrsi} data can be beneficial for reaching the desired resolution. Iqbal et al. \cite{iqbal_super-resolution_2019} propose a supervised \ac{dl} method for super-resolution of \ac{mrsi} data. They propose a U-Net architecture to take a T1w image and corresponding \ac{lrsi} image as inputs and produce a \ac{hrsi} image as output. In total, three different U-Nets are trained for three different \ac{lrsi} resolutions (16x16, 24x24 and 32x32). They use synthetic data, generated by a \ac{mrsi} simulator, for training and testing. This generator produces T1w images and a pair of \ac{lrsi} and \ac{hrsi} data.
The synthetic data is used to evaluate the trained models based on the \ac{mse} of the \ac{hrsi} reconstructions and the reconstruction of individual spectra at different resolutions. Additionally, the models are tested on downsampled in-vivo \ac{hrsi} data. During evaluation, different noise levels are tested and a comparison is performed with standard methods like zero-filling and bicubic interpolation. Results show that the \ac{dl} method performs better than standard methods for all noise levels and can be used as a denoising or acceleration method. 

In a follow-up study, Dong et al. \cite{dong_high-resolution_2021} propose another super-resolution model which included in-vivo \ac{mri} and \ac{mrsi} data. A dataset of \ac{hrsi} (64$\times$64) data is acquired and down-sampled to obtain \ac{lrsi} (16$\times$16) data. In total, 320 metabolic maps from 3 different patients are used as \ac{gt}. The \ac{dl} method consists of a U-Net with four different input modalities: \ac{lrsi} data, T1-weighted images, \ac{flair} images and contrast-enhanced T1-weighted images. The decoding part of the network uses spatial attention modules which automatically calculate spatial weight maps to focus on important features in each input modality. The \ac{hrsi} output is used to calculate the loss function, which consists of three different parts. The first part is a pixelwise \ac{mse} loss, calculated by comparing the output with the \ac{gt} data. To account for inter-pixel correlations, a second term called the \ac{ms-ssim} is added. The \ac{ms-ssim} measures the similarity between two different images at multiple scales in terms of luminance, contrast and structure. Finally, a discriminator is added to calculate an adversarial loss and capture more complex features. This study investigates the contribution of all different loss terms and input modalities by training multiple models. Results show that including all aforementioned inputs and losses achieves the best performance, also out-performing standard bi-cubic interpolation.

\subsection{Frequency \& Phase Correction}
\Ac{fpc} is necessary to reduce the effects of scanner frequency drifts, subject motion, or other inconsistencies to ensure reliable quantification without line broadening or loss in \ac{snr}. This is especially crucial in J-difference edited \ac{mrs} spectra which rely on accurately subtracting two aligned spectra.

Tapper et al. \cite{tapper_frequency_2021} propose an automated \ac{fpc} framework consisting of two separate \acp{nn}. A simulated dataset set is obtained using ideal excitation/refocusing pulses and shaped editing pulses. The simulation parameters are chosen to match the in-vivo data (Big GABA repository~\cite{mikkelsen_big_2017}) as close as possible.
Both corrections are implemented with two sequentially placed fully connected \acp{nn}, where the frequency shifts are estimated first, followed by an estimation of the phase offsets. Both models are trained individually taking the magnitude or real-only spectra, respectively, as the input to predict the frequency or phase offsets. Results for simulated test data show accurate predictions with a mean frequency offset error reported at $0.00 \pm 0.03$ Hz and a mean phase offset error of $-0.11 \pm 0.25$ degrees. Evaluation with in-vivo data shows similar performance to a model-based \ac{sr} method \cite{near_frequency_2015}, however, they indicate a substantial performance degradation compared to the simulated data.

Based on this work, Ma et al. \cite{ma_mr_2022} propose an alternative \ac{dl} approach for \ac{fpc} using \ac{cnn} architectures. They use identical parameter configurations to simulate MEGA-\ac{press} data for training and validation. However, they create additional spectra with lower \acp{snr} by adding Gaussian noise. The authors observe less subtraction artifacts with the \ac{cnn} and show an overall better performance than the approach of Tapper et al.\cite{tapper_frequency_2021}. For the simulated data the authors report a mean frequency offset error of $0.02 \pm 0.02$ Hz for the \ac{mlp} and $0.01 \pm 0.01$ Hz for the \ac{cnn}, and a mean phase offset error of $0.19 \pm 0.17$ degree for the \ac{mlp} and $0.12 \pm 0.09$ degree for the \ac{cnn}. For the in-vivo scenarios, performance is measured based on the variance of the choline metabolite of the spectra subtraction. The \ac{cnn} performs better in 66.67\% , 60.61\%, and 75.76\% of the 33 datasets, when small, medium, and large offsets are added to the data, respectively.

Shamaei at al. \cite{shamaei_modelinformed_2022} develop unsupervised methods for \ac{fpc} including two different \ac{cemd} models in which the input spectra are used as targets during training. One model focuses on reference peak fitting of \ac{cr} and uses a convolutional encoder to construct a lower dimensional latent representation of the input. The latent parameters are used as input for a Lorentzian lineshape model decoder. The second \ac{cemd} model uses the same encoder, but a \ac{sr} function is used as decoder. To account for unstable frequency components, both models are also trained for a limited frequency range (2.5 to 3.5 ppm). Training and validation is done on a simulated dataset with known frequency and phase offsets and evaluation is done on phantom data and in-vivo data from the Big GABA repository. The proposed unsupervised methods are compared with commonly used \ac{fpc} methods (\ac{sr} and \ac{cr} referencing, both on full and limited frequency ranges) and previous \ac{dl} methods~\cite{tapper_frequency_2021, ma_mr_2022}. In contrast to the non-\ac{dl} \ac{fpc} methods, the \ac{cemd} models perform equally well when trained and applied on a limited frequency range and their performance is less influenced by the presence of nuisance peaks. Compared to the previous \ac{dl} methods, the \ac{cemd} is able to train in an unsupervised fashion and requires only one network for both frequency and phase correction.   

\subsection{Ghosting Artifact Removal}
Ghosting artifacts, or so-called spurious echos, are usually caused by insufficient spoiling gradient power in combination with local susceptibility variations. These artifacts negatively influence the reliability of metabolite quantification as they may overlap with metabolite peaks.

Kyathanahally et al. \cite{kyathanahally_deep_2018} evaluate two networks based on fully connected \acp{nn} and \acp{cnn} for the detection as well as an autoencoder network with residual blocks for the removal of ghosting artifacts. Brain metabolite spectra are simulated for ideal \ac{press} characteristics, and ghosting artifacts with varying line-widths and amplitudes are added randomly. The fully connected \ac{nn} is trained on \ac{1d} spectra as input and class probabilities as output. An alternative classification approach is implemented with a \ac{cnn}, taking the real and imaginary part of a \ac{2d} spectrogram as input. The spectrograms are obtained by segmenting the time domain signals followed by a Fourier transform of each segment, creating \ac{2d} time-frequency spectrograms. To effectively remove the ghosting artifacts, an autoencoder network is implemented, taking the \ac{2d} corrupted spectrograms as input and outputting artifact-free spectrograms. The authors report a classification accuracy between 50\% and 75\% for the fully connected \ac{nn} depending on the number of layers. In contrast, the \ac{cnn} approach shows promising performance with a mean accuracy of 94\% for smaller datasets and an accuracy of over 99\% for larger training sets and subsequent in-vivo evaluations. In addition, the autoencoder method is found to be effective in removing ghost artifacts from distorted spectra, with low \ac{rmse} reported for the difference between ground truth and restored spectra in simulated data. However, the restoration is found to be suboptimal for in-vivo cases. 

\subsection{General Artifact Removal}
The previously discussed methods all focus on a specific processing step, yet \acp{nn} have the ability to learn more complex mappings from training data, enabling multiple artifact corrections at once. Lee and Kim \cite{lee_intact_2019} propose a \ac{cnn} architecture taking spectra contaminated with artifacts as input and predicting noise-free, metabolite-only spectra. The \ac{nn} is trained using simulated data with metabolite phantom spectra as a basis set and knowledge of in-vivo data for specific linewidth, \ac{snr}, and baseline ranges. Further, in-vivo test data is obtained from five healthy volunteers with identical scanner settings used for the phantom spectra of the basis set. For both simulated and in-vivo data, the \ac{cnn} clearly manages to obtain spectra with removed noise, narrow linewidth, removed frequency and phase shifts, and without spectral baseline. The results show no visible residual signal and thus suggest good removal of artifacts only in the simulated scenario. Further, the reported \ac{mape} of the quantification estimates and the \acp{gt} concentrations is $20.67\% \pm 16.71\%$. For the in-vivo spectra the results of the proposed method coincide with estimates of LCModel as well as commonly reported metabolite concentrations from the literature.

\section{Analysis}\label{sec:analysis}
The \ac{mrs} analysis focuses on converting the (processed) signals into meaningful and reliable metabolite concentration estimates. The following subsections summarize recent \ac{ml} applications for spectral quality assurance, metabolite quantification, uncertainty measurement, and classification.

\subsection{Quality Assurance}
\Ac{mrs} and \ac{mrsi} methods are susceptible to various imperfections causing artifacts in the acquired spectra which in turn can create unreliable and inaccurate measurements. This is limiting the clinical use of \ac{mrs} and causes a dependence on technical experts to inspect the spectral quality.

De Barros et al. \cite{pedrosa_de_barros_improving_2017} introduce an active learning method to improve labeling efficiency based on spectral quality by either accepting or rejecting spectra for further analysis. The method uses a dataset of over 28,000 in-vivo spectra from brain tumor patients. Forty-seven features are extracted from the time-domain and frequency-domain magnitude spectra and are used as input for a \ac{rf} classifier. Two expert spectroscopists manually label the spectra to provide \ac{gt} for supervised training. The active learning strategy employs an uncertainty range defined as $[0.5-\alpha, 0.5 + \alpha]$, where $\alpha$ controls the width. When the \ac{rf} classifier's output falls within this range, the uncertain data instance is added to the training set. This active learning method is evaluated iteratively, where in each iteration spectra from one patient are evaluated and uncertain examples are added to the training set. After retraining, the \ac{rf} classifier is validated on one patient (leave-one-out-cross-validation). Results show insignificant or minor differences in classification performance between different values for $\alpha$ ($0.1 \leq \alpha \leq 0.5$), resulting in an efficient way of training a \ac{rf} classifier with fewer manual labeling.    

Gurbani et al. \cite{gurbani_convolutional_2018} propose a \ac{cnn} architecture to automatically classify the quality of a given spectrum and integrate it into a software pipeline enabling real-time filtering of \ac{epsi} data. In-vivo spectra from patients with glioblastoma are collected after appropriate filtering. These spectra are then reviewed by \ac{mrs} experts and classified as "Good", "Acceptable", or "Poor" quality to obtain \ac{gt} labels for the \ac{nn}. The \ac{cnn} architecture takes the normalized real component of the spectrum as input and splits it into six specific regions, each with its own designated \ac{cnn}. These \acp{cnn} are trained in parallel by passing their concatenated outputs through a \ac{mlp} of which the output represents the probability to be classified as "Good". The overall model performs well in terms of detecting "Poor" quality spectra with an \ac{auc} of 0.951.  

Kyathanahally et al. \cite{kyathanahally_quality_2018} evaluate various \ac{ml} approaches for fast quality classification of \ac{mr} spectra. The authors perform training and testing on the multi-center studies INTERPRET~\cite{perez-ruiz_interpret_2010} and eTUMOUR~\cite{julia-sape_strategies_2012} consisting of more than 1000 spectra, mostly already classified into good and bad quality. Furthermore, they create an intermediate class for "Poor" quality if one of the three experts thought the spectrum was acceptable, and they also create their own local expert ratings for some previously unlabeled data. The authors evaluate various classifiers based on \acp{svm}, \ac{lda}, and \ac{rusboost} in combination with \ac{ica} and \ac{pca} or \ac{sffs} and \ac{tree} as feature extraction or feature selection methods. Their final approach uses a high number of features as input and a \ac{rusboost} classifier that undersamples to combat the imbalanced training data, showing a comparable performance in rejecting unsuitable spectra to a human expert.

Hern\'andez‐Villegas et al. \cite{hernandez-villegas_extraction_2022} propose a \ac{cnmf} for the same multi-center studies used in \cite{kyathanahally_quality_2018} to distinguish between good and poor quality spectra. The method first iteratively factorizes observations into a source matrix (of data centroids) and a mixing matrix (containing combination weights). Then, two experts define quality measures based on correlation and Euclidean distance of the extracted sources of 10 repetitions as well as based on the coding coefficient of the mixing matrices. Thereby, spectral quality can be assessed and characterized in an unsupervised fashion. The obtained results indicate that the defined quality measures can identify sources containing artifacts and the approach manages to distinguish between good and poor-quality spectra.

Jang et al. \cite{jang_unsupervised_2021} train a \ac{gan} to detect abnormalities in 3T human brain spectra. Normal and abnormal brain spectra are simulated, similarly to Lee and Kim \cite{lee_intact_2019}. Eight different classes of spectra are generated that are abnormal in \ac{snr}, linewidth, a single metabolite concentration, multiple metabolites concentrations (9), or all factors combined. Additionally, spectra that contain ghosting, residual water, or residual lipid artifacts are simulated with the help of phantom data. After training the \ac{gan} on normal spectra only, latent space mapping is performed. This mapping is done with a loss function containing a dissimilarity term that compares the generator output with the input spectra, and a discriminator term that uses feature matching from the second last layer of the discriminator. The classification between normal and abnormal spectra is performed with a \ac{2d} threshold using \ac{nmse} and the standard deviation of the spectra. Results show over 80\% accuracy for some abnormalities such as \ac{snr} and \ac{naa} concentration. Additionally, the \ac{gan} also detects ghosting, residual water, and residual lipid artifacts without using those spectra in the training phase. However, the model cannot accurately detect abnormalities in linewidth and low-concentrated metabolites with accuracies of around 50\%.

\subsection{Quantification}
Quantification aims at converting processed \ac{mrs} spectra/\acp{fid} into specific metabolite concentration estimates. Traditionally, model-based methods employed for metabolite quantification include linear combination model fitting, peak fitting, and peak integration \cite{vanhamme_mr_2001, poullet_mrs_2008}.

An early \ac{ml}-based quantification approach is introduced by Das et al.~\cite{das_quantification_2017} where they propose a \ac{rf} regression method. The proposed model is developed with different combinations of artificial spectra and in-vivo spectra. The \ac{rf} consists of a set of binary trees with splits based on random subsets of the feature variables on which the forest is subsequently trained. The trees of the \ac{rf} are trained using piece-wise linear regression over the input spectrum outputting metabolite concentration estimates, followed by taking the weighted average of the predictions from each tree to obtain a single output estimate. Results show that the \ac{rf} technique has similar performance as LCModel and could therefore be used in combination with LCModel to assist with the quantification of noisy spectra and enable faster convergence.

Hatami et al. \cite{hatami_magnetic_2018} propose a supervised \ac{cnn} model for metabolite quantification which is able to cover 20 different metabolites and a macromolecule signal. The \ac{cnn} takes real and imaginary parts of the spectra as a two-channel input and outputs concentrations of all metabolites of interest. A training and test with \ac{gt} concentrations are obtained by simulating spectra based on an \ac{mrs} signal model. To evaluate the accuracy of the quantification model, the \ac{smape} over the whole test set is calculated and compared with the performance of the model-based QUEST \cite{ratiney_time-domain_2005} method and the previously mentioned \ac{rf} regression algorithm from Das et al.~\cite{das_quantification_2017}. The proposed \ac{cnn} outperforms the other methods with and without the addition of noise.

Shamaei et al. \cite{shamaei_wavelet_2021} investigate the use of a \ac{cnn} which used a wavelet scattering transformation to extract features from the \ac{mrs} signal. The extracted features are fed into a fully connected feed-forward \ac{nn} to predict the relative amplitudes of the metabolite basis spectra, which can be used for absolute quantification. For training and evaluation, multiple datasets are simulated by using a signal-based model and uniform sampling of its parameters. Their model shows better performance, in terms of \ac{smape}, compared to QUEST and similar performance as the model of Hatami et al. \cite{hatami_magnetic_2018}. Additionally, the wavelet scattering \ac{cnn} shows robustness against metabolite phase changes and nuisance signals, such as macromolecules. 

Gurbani et al. \cite{gurbani_incorporation_2019} use a \ac{cemd} for spectral fitting. This two-step unsupervised \ac{dl} approach takes the real part of the spectrum as an input. During the first encoder-decoder step, a spectrum is mapped to a lower-dimensional space and a baseline is reconstructed using a wavelet reconstruction decoder. The resulting baseline is subtracted from the input spectrum and fed into the second encoder-decoder network, which is used to reconstruct the spectral lineshape, and therefore the fitting parameters for the metabolites of interest. Three metabolites are considered including \ac{cho}, \ac{cr}, and \ac{naa}. The final \ac{cemd} model is also included in a pipeline for creating whole-brain metabolite maps for patients with glioblastoma and is compared with MIDAS~\cite{maudsley_mapping_2009}. Results show that both methods have similar fitting performance and the \ac{cho}/\ac{naa} maps created by the \ac{cemd} have a Dice score of 0.72 when compared to MIDAS.   

Lee and Kim \cite{lee_deep_2020} propose a \ac{cnn} architecture to quantify metabolite concentrations. Their approach consists of a designated \ac{cnn} per metabolite, taking the real part of a spectrum as input to estimate the corresponding metabolite spectrum. The actual quantification of the metabolites is then obtained by computing the areas of the known spectral regions relative to the methyl signal of \ac{tcr}. The results for the proposed quantification approach show a well-performing algorithm for the completely synthetic scenarios (i.e. \ac{mape} of 1.92\% and 2.56\% for the methyl ($\sim3.0$ ppm) and methylene ($\sim3.9$ ppm) peaks for the reference metabolite \ac{tcr}). For the simulated spectra using metabolite phantoms and in-vivo baselines the mean \ac{mape} is increased as depicted above and ranged from $14.79 \pm 11.12\%$ to $23.07 \pm 16.36\%$ over the major metabolites.

Similarly, Iqbal et al. \cite{iqbal_deep_2021} propose a method with a designated \ac{nn} per metabolite implemented using U-Nets. Their approach takes the real, imaginary, and magnitude information of a fully sampled \ac{lcosy} spectrum as input and outputs the magnitude spectrum for each of the seventeen metabolites. The authors observe an increase in error for both degrading \ac{snr} and higher water signal amplitude. Nonetheless, the model shows accurate quantification of the metabolites, even for low concentrations. 

Rizzo et al. \cite{rizzo_quantification_2023} compare \ac{mrs} quantification using various \ac{cnn} models, input types, and learning methods. They use a simulated artifact-free dataset with \ac{gt} concentrations to enable fair comparison with standard model fitting. Results indicate that \ac{2d} spectrogram inputs outperform \ac{1d} frequency domain inputs and that including a water reference peak improves performance. The best model is a heterogeneous ensemble combining \ac{1d} and \ac{2d} inputs while increasing dataset size and applying active learning strategies do not significantly improve performance. However, \ac{dl}-based quantification still underperforms compared to standard model fitting and is highly biased toward training data when \ac{snr} is low. 

Schmid et al. \cite{schmid_deconvolution_2023} propose a \ac{dl}-based peak detection method as part of classical peak fitting. A simulated dataset, including various distortions, is used for training. Their \ac{cnn} model with \ac{lstm} blocks outputs classes (baseline, narrow peak, or broad peak) and values for the peak widths. Input spectra are dynamically scaled to enhance local contrast and peak labels are acquired with an automatic labeling procedure. The outputs are used to fix the number of peaks and initialize the peak width values for a classical peak fitting algorithm. Evaluations on simulated and experimental data show high scores on picking accuracy, spectral reconstruction, and sparsity of the peak selection. Their method outperforms using manual peak picking in terms of \ac{mae}, especially in crowded regions (i.e. $82$\% lower \ac{mae}). The authors state that their method, although optimized for high-field proton spectroscopy, is adaptable to different domains.     

Shamaei at al. \cite{shamaei_physics-informed_2023} implements a physics-informed \ac{dl} method to quantify simulated spectra and in-vivo spectra from the Big GABA repository \cite{mikkelsen_big_2017}. They use a \ac{cemd} architecture with an encoder that outputs parameters for the signal-based model decoder. This decoder uses a metabolite basis set and a numeric, parameterized, or regulated parameterized \ac{mm} signal contribution as prior knowledge. Their experiments include the investigation of different architectures for the encoder, the use of different \ac{mm} models, and different dataset sizes. Results show comparable performance to traditional quantification methods and the ability to use this \ac{dl} approach for in-vivo data, with best performances for shallow \ac{cnn} encoders and minimum dataset sizes of 12,000 samples. Additionally, a numerical \ac{mm} signal is favorable above parameterized and regulated parameterized \ac{mm} models. Due to the unsupervised training approach and the significant reduction in computation time, this method could be used as a faster alternative to quantify large, in-vivo \ac{mrs} datasets.    

\subsection{Uncertainty Measurement}
Uncertainty measurements of metabolite concentration estimates, such as the \ac{crlb} or the \ac{crlbp} are crucial for assuring reliable results, yet for data-driven methods such measures are generally biased and alternative metrics are difficult to validate.

Lee and Kim \cite{lee_deep_2020} propose a \ac{cnn}-based approach to quantify metabolite concentrations and simultaneously obtain an uncertainty estimate for the output spectra. The approach is developed with simulated rat brain spectra and is further evaluated using phantom data and in-vivo rat brain spectra. The authors obtain a measurement uncertainty with respect to \ac{snr}, linewidth, and \ac{sbr} by constructing an uncertainty measurement database from the training data. The \ac{snr} and linewidth are estimated from the input spectrum, specifically from \ac{tnaa}, while the \ac{sbr} of each metabolite is measured from the predicted metabolite spectra. Then a \ac{3d} space of the quantitative errors is computed and stored for each target metabolite as a function of the \ac{snr}, linewidth, and \ac{sbr}. The estimated quantification uncertainty of the proposed method is highly correlated with the actual errors obtained in a purely simulated scenario (i.e. $0.81 \pm 0.13; 0.88 \pm 0.09$ for 15 major metabolites). The correlation predicted error and \ac{gt} error for the simulated spectra using metabolite phantoms and in-vivo baselines are slightly lower (i.e. 0.7 or higher ($0.78 \pm 0.05$) and statistically significant for all 15 major metabolites).

In another work of Lee and Kim \cite{lee_bayesian_2022}, they propose an alternative \ac{cnn} architecture and training procedure to obtain both an estimate of metabolite concentrations and their uncertainties. Using \ac{mcdo}~\cite{gal_dropout_2016} and a variance leveraging loss function based on the log-likelihood~\cite{kendall_what_2017} both epistemic (model) and aleatoric (data) uncertainty estimates are obtained for each metabolite concentration. The \ac{cnn} is trained with simulated spectra, further tested with in-vivo data, and shows comparable performance to model-based alternatives such as LCModel. 

The work of Rizzo et al. \cite{wang_reliability_2022} investigates the reliability of \ac{dl}-based quantification and compared it to common model fitting methods. For that purpose, they design a \ac{cnn} taking spectrograms as input and outputting normalized metabolite concentration estimates. In a similar fashion to Lee et al. \cite{lee_bayesian_2022} the authors use \ac{mcdo} for epistemic (model) uncertainty and metrics based on bias and spread of the predicted concentration distribution for aleatoric (data) uncertainty information. The results indicate that the \ac{cnn}'s predictions tend towards the mean of the test data in cases with high uncertainty, indicating the model is biased. Meanwhile, model fitting methods show on average to be unbiased.

\subsection{Classification}\label{sec:analysis_classification}
Classification in \ac{mrs} and \ac{mrsi} data is important for clinical applications in terms of diagnosis and disease monitoring. Instead of using metabolite concentrations, \ac{ml} methods can be trained to perform direct classification.

In a multi-class pediatric brain tumor classification problem, Zarinabad et al. \cite{zarinabad_multiclass_2017} show that various \ac{ml} methods can distinguish between three different tumor classes. An unbalanced dataset (in-vivo, 1.5T) is used in combination with \ac{bsmote} (based on the \ac{smote} algorithm \cite{chawla_smote_2002}) to increase classification performance. The trained classifiers consist of a \ac{rf} classifier with an adaptive number of trees and four different AdaBoostM1 algorithms using different weak learners: naive Bayes, \ac{svm}, \ac{nn} and \ac{lda}. Classification is done by either using the full spectra or the metabolite concentrations quantified by TARQUIN~\cite{wilson_constrained_2011}. Oversampling the minority class with \ac{bsmote} results in better classification performances of the trained classifiers, both for concentrations and spectral inputs. The best balanced accuracy rates are $0.93$ and $0.90$ for concentrations and spectral inputs respectively, with different combinations of classifiers and oversampling rates possible.

A similar classification problem is investigated in a subsequent study from Zarinabad et al. \cite{zarinabad_application_2018}. However, the spectra are acquired on 3T scanners from four different hospitals. The tested classification algorithms are \ac{lda}, \ac{svm} and \ac{rf} approaches and \ac{bsmote} is used to account for the class imbalance. \Ac{pca} is performed on the metabolite profiles to extract four principal components which are used as input for the classification algorithms. The results show a maximum balanced accuracy of $0.86$ when using \ac{svm} as a classification method, which compares favorably with a previous 1.5T multi-center study \cite{vicente_accurate_2013}. 

For the same multiclass tumor classification, Zhao et al. \cite{zhao_metabolite_2022} propose to add metabolite selection. This study compares \ac{pca} and multiclass \ac{roc} as metabolite selection methods for training \ac{ml} classifiers: \ac{lda}, $k$-nearest neighbors, naive Bayes, \ac{nn} and \ac{svm}. The classification with three tumor classes is done for both 1.5T and 3T \ac{svs} data from multiple sites and oversampling for minority classes is done using the \ac{smote} algorithm. Final classification accuracy is determined using $k$-fold and leave-one-out cross-validation. Metabolite selection using multiclass \ac{roc} shows a higher accuracy compared to \ac{pca} with the highest balanced classification accuracy of 85\% for the 1.5T data with \ac{svm} and 75\% for the 3T data with \ac{lda}. A more transparent and explainable tool for diagnosis is obtained by selection of metabolites for \ac{ml}-based classification.   

Dikaios \cite{dikaios_deep_2021} trains three different \ac{ml} methods to differentiate metastasis from glioblastoma brain tumors. The models include \ac{svm}, \ac{mlp}, and \ac{cnn}. Different versions of the models with varying hyperparameters and/or layers are tested on four different datasets consisting of real GE spectra with additional noise, real Philips spectra with additional noise, synthetic GE spectra, and synthetic Philips spectra. A total of 12 models are trained using long TE, short TE, and concatenations of both versions of the spectra. Evaluation of the results, in terms of \ac{roc}-\ac{auc} and accuracy, shows the best performance for the \ac{1d} \ac{cnn} when using synthetic data and the concatenated echo times ($>90\%$ accuracy).

\section{Artificial Data Generation}\label{sec:artificial_dat_gen}
Accessing in-vivo \ac{mrs}/\ac{mrsi} data is limited due to time-consuming acquisitions, non-standardized methods \cite{lin_minimum_2021}, and privacy concerns. To overcome this limitation, artificial data generation is used for developing \ac{ml} applications. All previously discussed studies, containing artificial data, use non-\ac{ml} based generation methods like data augmentation and model-based simulation.
Data augmentation is a method that artificially increases the size and variety of the used dataset by applying transformations to real samples. On the other hand, model-based simulation involves sampling the parameters of a parametric model to generate \ac{mrs} data, which is essentially an inverse use of signal-based fitting models. While the concepts of artificial data generation are the same, the exact implementation varies a lot per study. 

\Ac{ml} methods can also be applied for artificial data generation itself. The work of Olliverre et al. \cite{olliverre_generating_2018} focuses on generative models for creating \ac{mrs} spectra. This study compares three different models (\ac{gan}, \ac{dcgan}, and \ac{pmm} \cite{olliverre_pairwise_2017}) on their ability to generate \ac{mrs} spectra for three classes: healthy, low-grade and high-grade tissue. The models are trained on a dataset consisting of 137, 1.5T \ac{press} acquired in-vivo spectra. The \ac{gan} model uses a generator and discriminator with fully connected layers and the \ac{dcgan} uses a deeper architecture, which generally requires more data. The \ac{dcgan} is therefore trained by using the full training dataset as batch size with the addition of batch normalization to deal with the relatively small dataset. All models are trained on all three tissue types separately and the quality of the artificially generated data is tested by training a \ac{rf} classifier. Results show that datasets generated by the \ac{gan} and \ac{pmm} are able to train a \ac{rf} classifier to the same level as using real \ac{mrs} data. The \ac{dcgan} generated data has lower performance due to the small dataset size that underutilizes the potential of deep learning.

\section{Conclusion \& Outlook}\label{sec:conclusion_outlook}
This review highlights recent \ac{ml} studies within the field of proton \ac{mrs}, focusing primarily on processing and analysis of \ac{mrs} spectra and \ac{mrsi} images with less attention on data acquisition. Although some studies have applied \ac{ml} to volume selection for \ac{svs} and shimming, other aspects of data acquisition (e.g., pulse sequence design, suppression techniques, and excitation area) have so far been disregarded. Due to the hardware-dependent nature of such applications, they are not only more difficult to integrate, but also to develop and test. Meanwhile, topics like quality assurance, quantification, and classification are more commonly addressed. Through the ability to simulate processed spectra or to rely on large databases for training data, such \ac{ml} models are more straightforward to develop. In the future, more acquisition-oriented simulations could bridge this gap.

Recent \ac{ml} studies focus on \ac{dl} model types instead of classical \ac{ml} methods like \ac{svm}, \ac{rf}, \ac{lda}, or \ac{pca}. Table \ref{tab:overview_papers} reveals that the most commonly used model type is a \ac{cnn}. Additionally, most autoencoders, U-Nets, and ensemble model types also include convolutional layers. \Acp{cnn} are widely adapted in computer vision and medical imaging \cite{soffer_convolutional_2019, bhatt_cnn_2021} with their benefits of weight sharing, simultaneously extracting features and performing classification, and easy implementation into large-scale networks \cite{alzubaidi_review_2021}. While recent studies from Rizzo et al. \cite{rizzo_quantification_2023} and Shamaei et al. \cite{shamaei_physics-informed_2023} compare different model architectures, more comparison studies are still missing. Future work should focus on testing different and new model types like transformers that have shown potential in other medical imaging fields \cite{li_transforming_2023}. These types of studies can aid in finding the best practices for \ac{mrs} and \ac{mrsi} applications. As the field of \ac{ml} rapidly evolves, it is important to keep up-to-date with new developments and investigate their role in \ac{mrs} and other spectroscopy domains.      

Section \ref{sec:artificial_dat_gen} mentions that many studies use different techniques to generate artificial data to develop their \ac{ml} methods, making the comparability between different studies very challenging. Also, significant performance drops are observed when models are trained with artificial data and tested on in-vivo data, showing difficulties in transferability from artificial data to in-vivo data. This stems not only from lacking generalization capabilities of the \ac{ml} methods, but also from the difficulty of accurately replicating in-vivo data through simulation or synthesis. While metabolite signals are well understood using density matrix simulations, other signal contributions, like macromolecules, water/fat residuals, and other artifacts, remain challenging to simulate. Therefore, efforts that investigate and standardize (artificial) data generation, as well as augmentation techniques, are crucial for future research.

\Ac{ml} methods rely on their training data to learn meaningful tasks and are inherently biased towards this data. Without preventive or predictive measures there are no guarantees for the model's performance for inputs outside of this distribution \cite{carlini_towards_2017}. Furthermore, the model's output might even collapse to the mean of its target distribution for mismatched inputs \cite{carlini_towards_2017}. In a clinical setting, such behaviors need to be detected and removed. Reliable and broadly applicable uncertainty measures for \ac{ml} prediction are therefore crucial for or clinical applicability of \ac{ml} in \ac{mrs}. Additionally, deploying hybrid models (combined model-based and data-driven systems) can allow \ac{ml} contributions to be leveraged by physics-informed models that behave unbiased and have guarantees on their estimates \cite{shlezinger_model-based_2023}.

The clinical utility of \ac{ml} applications in \ac{mrs} and \ac{mrsi} is one of the most important aspects of this research field. Attempts to decrease human-expert involvement, decrease acquisition time, and increase robustness and generalizability of existing \ac{mrs} tools are therefore essential. \Ac{ml} methods should also be easy to interpret by clinicians to be useful in clinical workflows. While \ac{ml} in \ac{mrs} and \ac{mrsi} is still in the early stages, the discussed studies show great potential for clinical adoption with plenty of future research possibilities.

\section*{Acknowledgments}
This work was (partially) funded by Spectralligence (EUREKA IA Call, ITEA4 project 20209).

\subsection*{Author contributions}
Dennis M. J. van de Sande and Julian P. Merkofer contributed equally to this work.

\subsection*{Conflict of interest}
The authors declare no potential conflict of interest.

\bibliography{refs}%

\begin{thebibliography}{10}

\bibitem{buonocore_magnetic_2015}
Buonocore Michael~H., Maddock Richard~J.. Magnetic Resonance Spectroscopy of
  the Brain: A Review of Physical Principles and Technical Methods.  {\it
  Reviews in the Neurosciences. }2015;26(6):609--632.

\bibitem{faghihi_magnetic_2017}
Faghihi Reza, {Zeinali-Rafsanjani} Banafsheh, {Mosleh-Shirazi} Mohammad-Amin,
  et al. Magnetic {{Resonance Spectroscopy}} and Its {{Clinical Applications}}:
  {{A Review}}.  {\it Journal of Medical Imaging and Radiation Sciences.
  }2017;48(3):233--253.

\bibitem{wilson_methodological_2019}
Wilson Martin, Andronesi Ovidiu, Barker Peter~B., et al. Methodological
  Consensus on Clinical Proton {{MRS}} of the Brain: {{Review}} and
  Recommendations.  {\it Magnetic Resonance in Medicine. }2019;82(2):527--550.

\bibitem{landheer_theoretical_2020}
Landheer Karl, Schulte Rolf~F., Treacy Michael~S., Swanberg Kelley~M., Juchem
  Christoph. Theoretical Description of Modern {{1H}} in {{Vivo}} Magnetic
  Resonance Spectroscopic Pulse Sequences.  {\it Journal of Magnetic Resonance
  Imaging. }2020;51(4):1008--1029.

\bibitem{juchem_b0_2021}
Juchem Christoph, Cudalbu Cristina, {de Graaf} Robin~A., et al. B0 Shimming for
  in Vivo Magnetic Resonance Spectroscopy: {{Experts}}' Consensus
  Recommendations.  {\it NMR in Biomedicine. }2021;34(5):e4350.

\bibitem{jansen_1h_2006}
Jansen Jacobus F.~A., Backes Walter~H., Nicolay Klaas, Kooi M.~Eline. {{1H MR
  Spectroscopy}} of the {{Brain}}: {{Absolute Quantification}} of
  {{Metabolites}}.  {\it Radiology. }2006;240(2):318--332.

\bibitem{near_preprocessing_2021}
Near Jamie, Harris Ashley~D., Juchem Christoph, et al. Preprocessing, Analysis
  and Quantification in Single-Voxel Magnetic Resonance Spectroscopy: Experts'
  Consensus Recommendations.  {\it NMR in Biomedicine. }2021;34(5):e4257.

\bibitem{knoll_deep-learning_2020}
Knoll Florian, Hammernik Kerstin, Zhang Chi, et al. Deep-{{Learning Methods}}
  for {{Parallel Magnetic Resonance Imaging Reconstruction}}: {{A Survey}} of
  the {{Current Approaches}}, {{Trends}}, and {{Issues}}.  {\it IEEE Signal
  Processing Magazine. }2020;37(1):128--140.

\bibitem{montalt-tordera_machine_2021}
{Montalt-Tordera} Javier, Muthurangu Vivek, Hauptmann Andreas, Steeden
  Jennifer~Anne. Machine Learning in {{Magnetic Resonance Imaging}}: {{Image}}
  Reconstruction.  {\it Physica Medica. }2021;83:79--87.

\bibitem{pal_review_2022}
Pal Arghya, Rathi Yogesh. A Review and Experimental Evaluation of Deep Learning
  Methods for {{MRI}} Reconstruction.  {\it The Journal of Machine Learning for
  Biomedical Imaging. }2022;1:001.

\bibitem{zhu_applications_2019}
Zhu Guangming, Jiang Bin, Tong Liz, Xie Yuan, Zaharchuk Greg, Wintermark Max.
  Applications of {{Deep Learning}} to {{Neuro-Imaging Techniques}}.  {\it
  Frontiers in Neurology. }2019;10:869.

\bibitem{mohammed_review_2020}
Mohammed Bakhtyar~Ahmed, {Al-Ani} Muzhir~Shaban. Review {{Research}} of
  {{Medical Image Analysis Using Deep Learning}}.  {\it UHD Journal of Science
  and Technology. }2020;4(2):75--90.

\bibitem{uddin_progress_2017}
Uddin L~Q, Dajani D~R, Voorhies W, Bednarz H, Kana R~K. Progress and Roadblocks
  in the Search for Brain-Based Biomarkers of Autism and
  Attention-Deficit/Hyperactivity Disorder.  {\it Translational Psychiatry.
  }2017;7(8):e1218-e1218.

\bibitem{rashid_towards_2020}
Rashid Barnaly, Calhoun Vince. Towards a Brain-based Predictome of Mental
  Illness.  {\it Human Brain Mapping. }2020;41(12):3468--3535.

\bibitem{pilmeyer_functional_2022}
Pilmeyer Jesper, Huijbers Willem, Lamerichs Rolf, Jansen Jacobus F.~A.,
  Breeuwer Marcel, Zinger Svitlana. Functional {{MRI}} in Major Depressive
  Disorder: {{A}} Review of Findings, Limitations, and Future Prospects.  {\it
  Journal of Neuroimaging. }2022;32(4):582--595.

\bibitem{santana_rs-fmri_2022}
Santana Caio~Pinheiro, {de Carvalho} Emerson~Assis, Rodrigues Igor~Duarte,
  Bastos Guilherme~Sousa, {de Souza} Adler~Diniz, {de Brito} Lucelmo~Lacerda.
  Rs-{{fMRI}} and Machine Learning for {{ASD}} Diagnosis: A Systematic Review
  and Meta-Analysis.  {\it Scientific Reports. }2022;12(1):6030.

\bibitem{chen_review_2020}
Chen Dicheng, Wang Zi, Guo Di, Orekhov Vladislav, Qu~Xiaobo. Review and
  {{Prospect}}: {{Deep Learning}} in {{Nuclear Magnetic Resonance
  Spectroscopy}}.  {\it Chemistry \textendash{} A European Journal.
  }2020;26(46):10391--10401.

\bibitem{rajeev_review_2021}
Rajeev S.K., Rajasekaran M.~Pallikonda, Krishna~Priya R., Al~Bimani Ali. A
  {{Review}} on {{Magnetic Resonance Spectroscopy}} for {{Clinical Diagnosis}}
  of {{Brain Tumour}} Using {{Deep Learning}}.  In: :461--465; 2021.

\bibitem{wang_recent_2020}
Wang Xizhao, Zhao Yanxia, Pourpanah Farhad. Recent Advances in Deep Learning.
  {\it International Journal of Machine Learning and Cybernetics.
  }2020;11(4):747--750.

\bibitem{marcus_next_2020}
Marcus Gary. {\it The {{Next Decade}} in {{AI}}: {{Four Steps Towards Robust
  Artificial Intelligence}}. } 2020.

\bibitem{sarker_deep_2021}
Sarker Iqbal~H.. Deep {{Learning}}: {{A Comprehensive Overview}} on
  {{Techniques}}, {{Taxonomy}}, {{Applications}} and {{Research Directions}}.
  {\it SN Computer Science. }2021;2(6):420.

\bibitem{bolan_automated_2020}
Bolan Patrick~J., Branzoli Francesca, Di~Stefano Anna~Luisa, et al. Automated
  {{Acquisition Planning}} for {{Magnetic Resonance Spectroscopy}} in {{Brain
  Cancer}}.  In:  Martel Anne~L., Abolmaesumi Purang, Stoyanov Danail, et al. ,
  eds. {\it Medical {{Image Computing}} and {{Computer Assisted Intervention}}
  \textendash{} {{MICCAI}} 2020}, Lecture {{Notes}} in {{Computer
  Science}}:730--739{Springer International Publishing}; 2020; {Cham}.

\bibitem{becker_deep_2022}
Becker Moritz, Jouda Mazin, Kolchinskaya Anastasiya, Korvink Jan~G.. Deep
  Regression with Ensembles Enables Fast, First-Order Shimming in Low-Field
  {{NMR}}.  {\it Journal of Magnetic Resonance. }2022;336:107151.

\bibitem{becker_acquisitions_2022}
Becker Moritz, Lehmkuhl S{\"o}ren, Kesselheim Stefan, Korvink Jan~G., Jouda
  Mazin. Acquisitions with Random Shim Values Enhance {{AI-driven NMR}}
  Shimming.  {\it Journal of Magnetic Resonance. }2022;345:107323.

\bibitem{nassirpour_multinet_2018}
Nassirpour Sahar, Chang Paul, Henning Anke. {{MultiNet PyGRAPPA}}: {{Multiple}}
  Neural Networks for Reconstructing Variable Density {{GRAPPA}} (a {{1H FID
  MRSI}} Study).  {\it NeuroImage. }2018;183:336--345.

\bibitem{lee_reconstruction_2020}
Lee Hyochul, Lee Hyeong~Hun, Kim Hyeonjin. Reconstruction of Spectra from
  Truncated Free Induction Decays by Deep Learning in Proton Magnetic Resonance
  Spectroscopy.  {\it Magnetic Resonance in Medicine. }2020;84(2):559--568.

\bibitem{luo_fast_2020}
Luo Jie, Zeng Qing, Wu~Ke, Lin Yanqin. Fast Reconstruction of Non-Uniform
  Sampling Multidimensional {{NMR}} Spectroscopy via a Deep Neural Network.
  {\it Journal of Magnetic Resonance. }2020;317:106772.

\bibitem{iqbal_deep_2021}
Iqbal Zohaib, Nguyen Dan, Thomas Michael~Albert, Jiang Steve. Deep Learning Can
  Accelerate and Quantify Simulated Localized Correlated Spectroscopy.  {\it
  Scientific Reports. }2021;11(1):8727.

\bibitem{motyka_kspacebased_2021}
Motyka Stanislav, Hingerl Lukas, Strasser Bernhard, et al. k-{{Space}}-based
  Coil Combination via Geometric Deep Learning for Reconstruction of
  non-{{Cartesian MRSI}} Data.  {\it Magnetic Resonance in Medicine.
  }2021;86(5):2353--2367.

\bibitem{chan_improved_2022}
Chan Kimberly~L., Ziegs Theresia, Henning Anke. Improved Signal-to-noise
  Performance of {{MultiNet GRAPPA 1H FID MRSI}} Reconstruction with
  Semi-synthetic Calibration Data.  {\it Magnetic Resonance in Medicine.
  }2022;88(4):1500--1515.

\bibitem{lei_deep_2021}
Lei Yang, Ji~Bing, Liu Tian, Curran Walter~J., Mao Hui, Yang Xiaofeng. Deep
  Learning-Based Denoising for Magnetic Resonance Spectroscopy Signals.  In:
  Gimi Barjor~S., Krol Andrzej, eds. {\it Medical {{Imaging}} 2021:
  {{Biomedical Applications}} in {{Molecular}}, {{Structural}}, and
  {{Functional Imaging}}}, :3{SPIE}; 2021; {Online Only, United States}.

\bibitem{iqbal_super-resolution_2019}
Iqbal Zohaib, Nguyen Dan, Hangel Gilbert, Motyka Stanislav, Bogner Wolfgang,
  Jiang Steve. Super-{{Resolution 1H Magnetic Resonance Spectroscopic Imaging
  Utilizing Deep Learning}}.  {\it Frontiers in Oncology. }2019;9:1010.

\bibitem{dong_high-resolution_2021}
Dong Siyuan, Hangel Gilbert, Bogner Wolfgang, et al. High-{{Resolution Magnetic
  Resonance Spectroscopic Imaging}} Using a {{Multi-Encoder Attention U-Net}}
  with {{Structural}} and {{Adversarial Loss}}.  In: :2891--2895{IEEE}; 2021;
  {Mexico}.

\bibitem{tapper_frequency_2021}
Tapper Sofie, Mikkelsen Mark, Dewey Blake~E., et al. Frequency and Phase
  Correction of {{J}}-difference Edited {{MR}} Spectra Using Deep Learning.
  {\it Magnetic Resonance in Medicine. }2021;85(4):1755--1765.

\bibitem{ma_mr_2022}
Ma~David~J., Le~Hortense A-M., Ye~Yuming, et al. {{MR}} Spectroscopy Frequency
  and Phase Correction Using Convolutional Neural Networks.  {\it Magnetic
  Resonance in Medicine. }2022;87(4):1700--1710.

\bibitem{shamaei_modelinformed_2022}
Shamaei Amirmohammad, Starcukova Jana, Pavlova Iveta, Starcuk Zenon.
  Model-informed Unsupervised Deep Learning Approaches to Frequency and Phase
  Correction of {{MRS}} Signals.  {\it Magnetic Resonance in Medicine.
  }2022;:mrm.29498.

\bibitem{kyathanahally_deep_2018}
Kyathanahally Sreenath~P., D{\"o}ring Andr{\'e}, Kreis Roland. Deep Learning
  Approaches for Detection and Removal of Ghosting Artifacts in {{MR}}
  Spectroscopy: {{Detection}} and {{Removal}} of {{Ghosting Artifacts}} in
  {{MRS Using Deep Learning}}.  {\it Magnetic Resonance in Medicine.
  }2018;80(3):851--863.

\bibitem{lee_intact_2019}
Lee Hyeong~Hun, Kim Hyeonjin. Intact Metabolite Spectrum Mining by Deep
  Learning in Proton Magnetic Resonance Spectroscopy of the Brain.  {\it
  Magnetic Resonance in Medicine. }2019;82(1):33--48.

\bibitem{pedrosa_de_barros_improving_2017}
{Pedrosa de Barros} Nuno, McKinley Richard, Wiest Roland, Slotboom Johannes.
  Improving Labeling Efficiency in Automatic Quality Control of {{MRSI}} Data.
  {\it Magnetic Resonance in Medicine. }2017;78(6):2399--2405.

\bibitem{gurbani_convolutional_2018}
Gurbani Saumya~S., Schreibmann Eduard, Maudsley Andrew~A., et al. A
  Convolutional Neural Network to Filter Artifacts in Spectroscopic {{MRI}}.
  {\it Magnetic Resonance in Medicine. }2018;80(5):1765--1775.

\bibitem{kyathanahally_quality_2018}
Kyathanahally Sreenath~P., Mocioiu Victor, {Pedrosa de Barros} Nuno, et al.
  Quality of Clinical Brain Tumor {{MR}} Spectra Judged by Humans and Machine
  Learning Tools.  {\it Magnetic Resonance in Medicine. }2018;79(5):2500--2510.

\bibitem{jang_unsupervised_2021}
Jang Joon, Lee Hyeong~Hun, Park Ji-Ae, Kim Hyeonjin. Unsupervised Anomaly
  Detection Using Generative Adversarial Networks in {{1H-MRS}} of the Brain.
  {\it Journal of Magnetic Resonance. }2021;325:106936.

\bibitem{hernandez-villegas_extraction_2022}
{Hern{\'a}ndez-Villegas} Yanisleydis, {Ortega-Martorell} Sandra, Ar{\'u}s
  Carles, Vellido Alfredo, {Juli{\`a}-Sap{\'e}} Margarida. Extraction of
  Artefactual {{MRS}} Patterns from a Large Database Using Non-Negative Matrix
  Factorization.  {\it NMR in Biomedicine. }2022;35(4):e4193.

\bibitem{das_quantification_2017}
Das Dhritiman, Coello Eduardo, Schulte Rolf~F., Menze Bjoern~H.. Quantification
  of {{Metabolites}} in {{Magnetic Resonance Spectroscopic Imaging Using
  Machine Learning}}.  In:  Descoteaux Maxime, {Maier-Hein} Lena, Franz Alfred,
  Jannin Pierre, Collins D.~Louis, Duchesne Simon, eds. {\it Medical {{Image
  Computing}} and {{Computer Assisted Intervention}} - {{MICCAI}} 2017},
  Lecture {{Notes}} in {{Computer Science}}:462--470{Springer International
  Publishing}; 2017; {Cham}.

\bibitem{hatami_magnetic_2018}
Hatami Nima, Sdika Micha{\"e}l, Ratiney H{\'e}l{\`e}ne. Magnetic {{Resonance
  Spectroscopy Quantification Using Deep Learning}}.  In:  Frangi Alejandro~F.,
  Schnabel Julia~A., Davatzikos Christos, {Alberola-L{\'o}pez} Carlos,
  Fichtinger Gabor, eds. {\it Medical {{Image Computing}} and {{Computer
  Assisted Intervention}} \textendash{} {{MICCAI}} 2018}, Lecture {{Notes}} in
  {{Computer Science}}:467--475{Springer International Publishing}; 2018;
  {Cham}.

\bibitem{gurbani_incorporation_2019}
Gurbani Saumya~S., Sheriff Sulaiman, Maudsley Andrew~A., Shim Hyunsuk, Cooper
  Lee~A.D.. Incorporation of a Spectral Model in a Convolutional Neural Network
  for Accelerated Spectral Fitting.  {\it Magnetic Resonance in Medicine.
  }2019;81(5):3346--3357.

\bibitem{lee_deep_2020}
Lee Hyeong~Hun, Kim Hyeonjin. Deep Learning-Based Target Metabolite Isolation
  and Big Data-Driven Measurement Uncertainty Estimation in Proton Magnetic
  Resonance Spectroscopy of the Brain.  {\it Magnetic Resonance in Medicine.
  }2020;84(4):1689--1706.

\bibitem{shamaei_wavelet_2021}
Shamaei Amirmohammad, Star{\v c}ukov{\'a} Jana, Star{\v c}uk~Jr. Zenon. A
  {{Wavelet Scattering Convolutional Network}} for {{Magnetic Resonance
  Spectroscopy Signal Quantitation}}:.  In: :268--275{SCITEPRESS - Science and
  Technology Publications}; 2021; {Online Streaming, --- Select a Country ---}.

\bibitem{rizzo_quantification_2023}
Rizzo Rudy, Dziadosz Martyna, Kyathanahally Sreenath~P., Shamaei Amirmohammad,
  Kreis Roland. Quantification of {{MR}} Spectra by Deep Learning in an
  Idealized Setting: {{Investigation}} of Forms of Input, Network
  Architectures, Optimization by Ensembles of Networks, and Training Bias.
  {\it Magnetic Resonance in Medicine. }2023;89(5):1707--1727.

\bibitem{schmid_deconvolution_2023}
Schmid N., Bruderer S., Paruzzo F., et al. Deconvolution of {{1D NMR}} Spectra:
  {{A}} Deep Learning-Based Approach.  {\it Journal of Magnetic Resonance.
  }2023;347:107357.

\bibitem{shamaei_physics-informed_2023}
Shamaei Amirmohammad, Starcukova Jana, Starcuk Zenon. Physics-Informed Deep
  Learning Approach to Quantification of Human Brain Metabolites from Magnetic
  Resonance Spectroscopy Data.  {\it Computers in Biology and Medicine.
  }2023;158:106837.

\bibitem{lee_bayesian_2022}
Lee Hyeong~Hun, Kim Hyeonjin. Bayesian Deep Learning\textendash Based
  {{1H}}-{{MRS}} of the Brain: {{Metabolite}} Quantification with Uncertainty
  Estimation Using {{Monte Carlo}} Dropout.  {\it Magnetic Resonance in
  Medicine. }2022;88(1):38--52.

\bibitem{wang_reliability_2022}
Rizzo Rudy, Dziadosz Martyna, Kyathanahally Sreenath~P., Reyes Mauricio, Kreis
  Roland. Reliability of {{Quantification Estimates}} in {{MR Spectroscopy}}:
  {{CNNs}} vs {{Traditional Model Fitting}}.  In:  Wang Linwei, Dou Qi,
  Fletcher P.~Thomas, Speidel Stefanie, Li~Shuo, eds. {\it Medical {{Image
  Computing}} and {{Computer Assisted Intervention}} \textendash{} {{MICCAI}}
  2022}, {Cham}: {Springer Nature Switzerland} 2022 (pp. 715--724).

\bibitem{zarinabad_multiclass_2017}
Zarinabad Niloufar, Wilson Martin, Gill Simrandip~K, Manias Karen~A, Davies
  Nigel~P, Peet Andrew~C. Multiclass Imbalance Learning: {{Improving}}
  Classification of Pediatric Brain Tumors from Magnetic Resonance
  Spectroscopy: {{Imbalanced}} Learning for {{MRS}} Tumor Classification.  {\it
  Magnetic Resonance in Medicine. }2017;77(6):2114--2124.

\bibitem{zarinabad_application_2018}
Zarinabad Niloufar, Abernethy Laurence~J., Avula Shivaram, et al. Application
  of Pattern Recognition Techniques for Classification of Pediatric Brain
  Tumors by in Vivo {{3T 1H-MR}} Spectroscopy -- {{A}} Multi-Center Study.
  {\it Magnetic Resonance in Medicine. }2018;79(4):2359--2366.

\bibitem{dikaios_deep_2021}
Dikaios Nikolaos. Deep Learning Magnetic Resonance Spectroscopy Fingerprints of
  Brain Tumours Using Quantum Mechanically Synthesised Data.  {\it NMR in
  Biomedicine. }2021;34(4):e4479.

\bibitem{zhao_metabolite_2022}
Zhao Dadi, Grist James~T., Rose Heather~E.L., et al. Metabolite Selection for
  Machine Learning in Childhood Brain Tumour Classification.  {\it NMR in
  Biomedicine. }2022;35(6):e4673.

\bibitem{olliverre_generating_2018}
Olliverre Nathan, Yang Guang, Slabaugh Gregory, {Reyes-Aldasoro}
  Constantino~Carlos, Alonso Eduardo. Generating {{Magnetic Resonance
  Spectroscopy Imaging Data}} of {{Brain Tumours}} from {{Linear}},
  {{Non-linear}} and~{{Deep Learning Models}}.  In:  Gooya Ali, Goksel Orcun,
  Oguz Ipek, Burgos Ninon, eds. {\it Simulation and {{Synthesis}} in {{Medical
  Imaging}}}, Lecture {{Notes}} in {{Computer Science}}:130--138{Springer
  International Publishing}; 2018; {Cham}.

\bibitem{kreis_terminology_2021}
Kreis Roland, Boer Vincent, Choi In-Young, et al. Terminology and Concepts for
  the Characterization of in Vivo {{MR}} Spectroscopy Methods and {{MR}}
  Spectra: {{Background}} and Experts' Consensus Recommendations.  {\it NMR in
  Biomedicine. }2021;34(5).

\bibitem{lecun_deep_2015}
LeCun Yann, Bengio Yoshua, Hinton Geoffrey. Deep Learning.  {\it Nature.
  }2015;521(7553):436--444.

\bibitem{goodfellow_deep_2016}
Goodfellow Ian, Bengio Yoshua, Courville Aaron. {\it Deep Learning}.
\newblock Adaptive Computation and Machine Learning{Cambridge, Massachusetts}:
  {The MIT Press}; 2016.

\bibitem{burkov_hundred-page_2019}
Burkov Andriy. {\it The Hundred-Page Machine Learning Book}.
\newblock {Polen}: {Andriy Burkov}; 2019.

\bibitem{de_graaf_vivo_2019}
De~Graaf Robin~A.. {\it In Vivo {{NMR}} Spectroscopy: Principles and
  Techniques}.
\newblock {Hoboken, NJ}: {John Wiley \& Sons, Inc}; 3rd ed~ed.2019.

\bibitem{press_numerical_2007}
Press William~H., ed.{\it Numerical Recipes: The Art of Scientific Computing}.
\newblock {Cambridge, UK ; New York}: {Cambridge University Press}; 3rd
  ed~ed.2007.

\bibitem{ronneberger_u-net_2015}
Ronneberger Olaf, Fischer Philipp, Brox Thomas. U-{{Net}}: {{Convolutional
  Networks}} for {{Biomedical Image Segmentation}}.  In:  Navab Nassir,
  Hornegger Joachim, Wells William~M., Frangi Alejandro~F., eds. {\it Medical
  {{Image Computing}} and {{Computer-Assisted Intervention}} \textendash{}
  {{MICCAI}} 2015}, Lecture {{Notes}} in {{Computer Science}}:234--241{Springer
  International Publishing}; 2015; {Cham}.

\bibitem{thomas_localized_2001}
Thomas M.~Albert, Yue Kenneth, Binesh Nader, et al. Localized Two-Dimensional
  Shift Correlated {{MR}} Spectroscopy of Human Brain.  {\it Magnetic Resonance
  in Medicine. }2001;46(1):58--67.

\bibitem{mayzel_causality_2014}
Mayzel M., Kazimierczuk K., Orekhov V.~{\relax Yu}.. The Causality Principle in
  the Reconstruction of Sparse {{NMR}} Spectra.  {\it Chem. Commun..
  }2014;50(64):8947--8950.

\bibitem{hyberts_application_2012}
Hyberts Sven~G., Milbradt Alexander~G., Wagner Andreas~B., Arthanari Haribabu,
  Wagner Gerhard. Application of Iterative Soft Thresholding for Fast
  Reconstruction of {{NMR}} Data Non-Uniformly Sampled with Multidimensional
  {{Poisson Gap}} Scheduling.  {\it Journal of Biomolecular NMR.
  }2012;52(4):315--327.

\bibitem{ying_sparse_2017}
Ying Jinfa, Delaglio Frank, Torchia Dennis~A., Bax Ad. Sparse Multidimensional
  Iterative Lineshape-Enhanced ({{SMILE}}) Reconstruction of Both Non-Uniformly
  Sampled and Conventional {{NMR}} Data.  {\it Journal of Biomolecular NMR.
  }2017;68(2):101--118.

\bibitem{vidya_shankar_fast_2019}
Vidya~Shankar Rohini, Chang John~C., Hu~Houchun~H., Kodibagkar Vikram~D.. Fast
  Data Acquisition Techniques in Magnetic Resonance Spectroscopic Imaging.
  {\it NMR in Biomedicine. }2019;32(3):e4046.

\bibitem{bogner_accelerated_2021}
Bogner Wolfgang, Otazo Ricardo, Henning Anke. Accelerated {{MR}} Spectroscopic
  Imaging\textemdash a Review of Current and Emerging Techniques.  {\it NMR in
  Biomedicine. }2021;34(5).

\bibitem{pruessmann_sense_1999}
Pruessmann K.~P., Weiger M., Scheidegger M.~B., Boesiger P.. {{SENSE}}:
  Sensitivity Encoding for Fast {{MRI}}.  {\it Magnetic Resonance in Medicine.
  }1999;42(5):952--962.

\bibitem{griswold_generalized_2002}
Griswold Mark~A., Jakob Peter~M., Heidemann Robin~M., et al. Generalized
  Autocalibrating Partially Parallel Acquisitions ({{GRAPPA}}).  {\it Magnetic
  Resonance in Medicine. }2002;47(6):1202--1210.

\bibitem{breuer_controlled_2006}
Breuer Felix~A., Blaimer Martin, Mueller Matthias~F., et al. Controlled
  Aliasing in Volumetric Parallel Imaging ({{2D CAIPIRINHA}}).  {\it Magnetic
  Resonance in Medicine. }2006;55(3):549--556.

\bibitem{zhou_graph_2020}
Zhou Jie, Cui Ganqu, Hu~Shengding, et al. Graph Neural Networks: {{A}} Review
  of Methods and Applications.  {\it AI Open. }2020;1:57--81.

\bibitem{moser_noncartesian_2019}
Moser Philipp, Bogner Wolfgang, Hingerl Lukas, et al. Non-{{Cartesian GRAPPA}}
  and Coil Combination Using Interleaved Calibration Data \textendash{}
  Application to Concentric-ring {{MRSI}} of the Human Brain at {{7T}}.  {\it
  Magnetic Resonance in Medicine. }2019;82(5):1587--1603.

\bibitem{provencher_estimation_1993}
Provencher Stephen~W.. Estimation of Metabolite Concentrations from Localizedin
  Vivo Proton {{NMR}} Spectra.  {\it Magnetic Resonance in Medicine.
  }1993;30(6):672--679.

\bibitem{mikkelsen_big_2017}
Mikkelsen Mark, Barker Peter~B., Bhattacharyya Pallab~K., et al. Big {{GABA}}:
  {{Edited MR}} Spectroscopy at 24 Research Sites.  {\it NeuroImage.
  }2017;159:32--45.

\bibitem{near_frequency_2015}
Near Jamie, Edden Richard, Evans C.~John, Paquin Rapha{\"e}l, Harris Ashley,
  Jezzard Peter. Frequency and Phase Drift Correction of Magnetic Resonance
  Spectroscopy Data by Spectral Registration in the Time Domain: {{MRS Drift
  Correction Using Spectral Registration}}.  {\it Magnetic Resonance in
  Medicine. }2015;73(1):44--50.

\bibitem{perez-ruiz_interpret_2010}
{P{\'e}rez-Ruiz} Alexander, {Juli{\`a}-Sap{\'e}} Margarida, Mercadal Guillem,
  Olier Iv{\'a}n, Maj{\'o}s Carles, Ar{\'u}s Carles. The {{INTERPRET
  Decision-Support System}} Version 3.0 for Evaluation of {{Magnetic Resonance
  Spectroscopy}} Data from Human Brain Tumours and Other Abnormal Brain Masses.
   {\it BMC Bioinformatics. }2010;11(1):581.

\bibitem{julia-sape_strategies_2012}
{Julia-Sape} M., Lurgi M., Mier M., et al. Strategies for Annotation and
  Curation of Translational Databases: The {{eTUMOUR}} Project.  {\it Database.
  }2012;2012(0):bas035-bas035.

\bibitem{vanhamme_mr_2001}
Vanhamme Leentie, Sundin Tomas, Hecke Paul~Van, Huffel Sabine~Van. {{MR}}
  Spectroscopy Quantitation: A Review of Time-Domain Methods.  {\it NMR in
  Biomedicine. }2001;14(4):233--246.

\bibitem{poullet_mrs_2008}
Poullet Jean-Baptiste, Sima Diana~M., Van~Huffel Sabine. {{MRS}} Signal
  Quantitation: {{A}} Review of Time- and Frequency-Domain Methods.  {\it
  Journal of Magnetic Resonance. }2008;195(2):134--144.

\bibitem{ratiney_time-domain_2005}
Ratiney H., Sdika M., Coenradie Y., Cavassila S., Ormondt D.,
  {Graveron-Demilly} D.. Time-Domain Semi-Parametric Estimation Based on a
  Metabolite Basis Set.  {\it NMR in Biomedicine. }2005;18(1):1--13.

\bibitem{maudsley_mapping_2009}
Maudsley A.A., Domenig C., Govind V., et al. Mapping of Brain Metabolite
  Distributions by Volumetric Proton {{MR}} Spectroscopic Imaging ({{MRSI}}).
  {\it Magnetic Resonance in Medicine. }2009;61(3):548--559.

\bibitem{gal_dropout_2016}
Gal Yarin, Ghahramani Zoubin. {\it Dropout as a {{Bayesian Approximation}}:
  {{Representing Model Uncertainty}} in {{Deep Learning}}. } 2016.

\bibitem{kendall_what_2017}
Kendall Alex, Gal Yarin. What {{Uncertainties Do We Need}} in {{Bayesian Deep
  Learning}} for {{Computer Vision}}?. 2017;.

\bibitem{chawla_smote_2002}
Chawla N.~V., Bowyer K.~W., Hall L.~O., Kegelmeyer W.~P.. {{SMOTE}}:
  {{Synthetic Minority Over-sampling Technique}}.  {\it Journal of Artificial
  Intelligence Research. }2002;16:321--357.

\bibitem{wilson_constrained_2011}
Wilson Martin, Reynolds Greg, Kauppinen Risto~A., Arvanitis Theodoros~N., Peet
  Andrew~C.. A Constrained Least-Squares Approach to the Automated Quantitation
  of in Vivo {\textsuperscript{1}} {{H}} Magnetic Resonance Spectroscopy Data:
  {{Automated Quantitation}} of {{In Vivo}} {\textsuperscript{1}} {{H MRS
  Data}}.  {\it Magnetic Resonance in Medicine. }2011;65(1):1--12.

\bibitem{vicente_accurate_2013}
Vicente Javier, {Fuster-Garcia} Elies, Tortajada Salvador, et al. Accurate
  Classification of Childhood Brain Tumours by in Vivo {$^{1}$}{{H MRS}} - a
  Multi-Centre Study.  {\it European Journal of Cancer (Oxford, England: 1990).
  }2013;49(3):658--667.

\bibitem{lin_minimum_2021}
Lin Alexander, Andronesi Ovidiu, Bogner Wolfgang, et al. Minimum {{Reporting
  Standards}} for in Vivo {{Magnetic Resonance Spectroscopy}} ({{MRSinMRS}}):
  {{Experts}}' Consensus Recommendations.  {\it NMR in Biomedicine.
  }2021;34(5).

\bibitem{olliverre_pairwise_2017}
Olliverre Nathan, Asad Muhammad, Yang Guang, Howe Franklyn, Slabaugh Gregory.
  Pairwise Mixture Model for Unmixing Partial Volume Effect in Multi-Voxel
  {{MR}} Spectroscopy of Brain Tumour Patients.  In: :449--461{SPIE}; 2017.

\bibitem{soffer_convolutional_2019}
Soffer Shelly, {Ben-Cohen} Avi, Shimon Orit, Amitai Michal~Marianne, Greenspan
  Hayit, Klang Eyal. Convolutional {{Neural Networks}} for {{Radiologic
  Images}}: {{A Radiologist}}'s {{Guide}}.  {\it Radiology.
  }2019;290(3):590--606.

\bibitem{bhatt_cnn_2021}
Bhatt Dulari, Patel Chirag, Talsania Hardik, et al. {{CNN Variants}} for
  {{Computer Vision}}: {{History}}, {{Architecture}}, {{Application}},
  {{Challenges}} and {{Future Scope}}.  {\it Electronics. }2021;10(20):2470.

\bibitem{alzubaidi_review_2021}
Alzubaidi Laith, Zhang Jinglan, Humaidi Amjad~J., et al. Review of Deep
  Learning: Concepts, {{CNN}} Architectures, Challenges, Applications, Future
  Directions.  {\it Journal of Big Data. }2021;8(1):53.

\bibitem{li_transforming_2023}
Li~Jun, Chen Junyu, Tang Yucheng, Wang Ce, Landman Bennett~A., Zhou S.~Kevin.
  Transforming Medical Imaging with {{Transformers}}? {{A}} Comparative Review
  of Key Properties, Current Progresses, and Future Perspectives.  {\it Medical
  Image Analysis. }2023;85:102762.

\bibitem{carlini_towards_2017}
Carlini Nicholas, Wagner David. Towards {{Evaluating}} the {{Robustness}} of
  {{Neural Networks}}.  In: :39--57; 2017.

\bibitem{shlezinger_model-based_2023}
Shlezinger Nir, Whang Jay, Eldar Yonina~C., Dimakis Alexandros~G..
  Model-{{Based Deep Learning}}.  {\it Proceedings of the IEEE. }2023;:1--35.

\end{thebibliography}


\end{document}